\documentclass[a4paper,fleqn,usenatbib]{mnras}

\usepackage{amsmath}	
\usepackage{amssymb}	
\usepackage{txfonts}

\usepackage[T1]{fontenc}
\usepackage{ae,aecompl}

\usepackage{graphicx}	

\usepackage{siunitx}

\defcitealias{DES15}{DES16}

\title[The Small Scales of Cosmic Shear]{Inference from the small scales of cosmic shear with current and future Dark Energy Survey data}

\usepackage{eso-pic}

\AddToShipoutPictureBG*{%
  \AtPageUpperLeft{%
    \hspace{0.75\paperwidth}%
    \raisebox{-3.5\baselineskip}{%
      \makebox[0pt][l]{\textnormal{DES-2016-0185}}
}}}%

\AddToShipoutPictureBG*{%
  \AtPageUpperLeft{%
    \hspace{0.75\paperwidth}%
    \raisebox{-4.5\baselineskip}{%
      \makebox[0pt][l]{\textnormal{FERMILAB-PUB-16-303-A}}
}}}%

\author[The DES Collaboration]{
\parbox{\textwidth}{
\Large
N.~MacCrann$^{1}$\thanks{E-mail: niall.maccrann@manchester.ac.uk},
J.~Aleksi\'c$^{2}$,
A.~Amara$^{3}$,
S.~L.~Bridle$^{1}$,
C.~Bruderer$^{3}$,
C.~Chang$^{3}$,
S.~Dodelson$^{4,5}$,
T.~F.~Eifler$^{6}$,
E.~M.~Huff$^{6}$,
D.~Huterer$^{7}$,
T.~Kacprzak$^{3}$,
A.~Refregier$^{3}$,
E.~Suchyta$^{8}$,
R.~H.~Wechsler$^{9,10,11}$,
J.~Zuntz$^{1}$,
T. M. C.~Abbott$^{12}$,
S.~Allam$^{4}$,
J.~Annis$^{4}$,
R.~Armstrong$^{13}$,
A.~Benoit-L{\'e}vy$^{14,15,16}$,
D.~Brooks$^{15}$,
D.~L.~Burke$^{10,11}$,
A. Carnero Rosell$^{17,18}$,
M.~Carrasco~Kind$^{19,20}$,
J.~Carretero$^{21,2}$,
F.~J.~Castander$^{21}$,
M.~Crocce$^{21}$,
C.~E.~Cunha$^{10}$,
L.~N.~da Costa$^{17,18}$,
S.~Desai$^{22,23}$,
H.~T.~Diehl$^{4}$,
J.~P.~Dietrich$^{22,23}$,
P.~Doel$^{15}$,
A.~E.~Evrard$^{24,7}$,
B.~Flaugher$^{4}$,
P.~Fosalba$^{21}$,
D.~W.~Gerdes$^{7}$,
D.~A.~Goldstein$^{25,26}$,
D.~Gruen$^{10,11}$,
R.~A.~Gruendl$^{19,20}$,
G.~Gutierrez$^{4}$,
K.~Honscheid$^{27,28}$,
D.~J.~James$^{12}$,
M.~Jarvis$^{8}$,
E.~Krause$^{10}$,
K.~Kuehn$^{29}$,
N.~Kuropatkin$^{4}$,
M.~Lima$^{30,17}$,
J.~L.~Marshall$^{31}$,
P.~Melchior$^{13}$,
F.~Menanteau$^{19,20}$,
R.~Miquel$^{32,2}$,
A.~A.~Plazas$^{6}$,
A.~K.~Romer$^{33}$,
E.~S.~Rykoff$^{10,11}$,
E.~Sanchez$^{34}$,
V.~Scarpine$^{4}$,
I.~Sevilla-Noarbe$^{34}$,
E.~Sheldon$^{35}$,
M.~Soares-Santos$^{4}$,
M.~E.~C.~Swanson$^{20}$,
G.~Tarle$^{7}$,
D.~Thomas$^{36}$,
V.~Vikram$^{37}$
\vspace{0.1cm}\\~\\
\begin{center} (The DES Collaboration) \\ \centering \textit{Author affiliations are listed at the end of this paper}\end{center}
}
}
\date{Accepted XXX. Received YYY; in original form ZZZ}

\pubyear{2016}

\newcommand{\be}{\begin{equation}}
\newcommand{\ee}{\end{equation}} 
\newcommand{\bal}{\begin{align}}
\newcommand{\eal}{\end{align}}
\newcommand{\bnum}{\begin{enumerate}}
\newcommand{\enum}{\end{enumerate}}

\newcommand{\ben}{\begin{enumerate}}
\newcommand{\bi}{\begin{itemize}}
\newcommand{\ei}{\end{itemize}}
\newcommand{\een}{\end{enumerate}}

\newcommand{\pofk}{P_{\delta}(k,z)}

\newcommand{\ngmix}{\textsc{ngmix}}

\newcommand{\sex}{\textsc{SExtractor}}
\newcommand{\halofit}{{\textsc{halofit}}}

\newcommand{\xip}{\xi_{+}}
\newcommand{\xim}{\xi_{-}}
\newcommand{\xipm}{\xi_{\pm}}
\newcommand*\mean[1]{\bar{#1}}

\newcommand{\dobs}{\delta_{\text{obs}}}
\newcommand{\dobsi}{\delta_{\text{obs},i}}
\newcommand{\dobsj}{\delta_{\text{obs},j}}
\newcommand{\kflux}{\kappa_{\text{flux}}}

\newcommand{\dx}[1]{\mathrm{d}{#1}\,}
\newcommand{\dsx}[1]{\mathrm{d}^2{#1}\,}

\newcommand\eqn[1]{equation~\ref{#1}}

\newcommand\fig[1]{Figure~\ref{#1}}

\newcommand\sect[1]{Section~\ref{#1}}
\newcommand{\snr}{\ensuremath{S/N}}

\DeclareSIUnit \h {\mbox{$h$}}
\DeclareSIUnit \degsq {\mbox{$\rm{deg}^2$}}

\DeclareSIUnit \parsec {pc}
\DeclareSIUnit \parsec {pc}
\DeclareSIUnit \megaparsec {Mpc}

\newcommand{\vect}[1]{\boldsymbol{\mathbf{#1}}}
\newcommand{\vectheta}{\boldsymbol{\thetaup}}
\newcommand{\obs}{\mathrm{obs}}

\newcommand{\planck}{\textit{Planck}}
\newcommand{\ctruth}{C_{\mathrm{truth}}}
\newcommand{\rtsase}{\sqrt{\sigma_{A}\sigma_{\eta_0}}}

\newcommand{\cosmosis}{\textsc{CosmoSIS}}

\begin{document}
\label{firstpage}
\pagerange{\pageref{firstpage}--\pageref{lastpage}}
\maketitle

\begin{abstract}
Cosmic shear is sensitive to fluctuations in the cosmological matter density field, including on small physical scales, where matter clustering is affected by 
baryonic physics in galaxies and galaxy clusters, such as star formation, supernovae feedback and AGN feedback. While muddying any cosmological information that is contained in small scale cosmic shear measurements, this does mean that cosmic shear has the potential to constrain baryonic physics and galaxy formation. We perform an analysis of the Dark Energy Survey (DES) Science Verification (SV) cosmic shear measurements, now extended to smaller scales, and using the \citet{mead15} halo model to account for baryonic feedback. While the SV data has limited statistical power, we demonstrate using a simulated likelihood analysis that the final DES data will have the statistical power to differentiate among baryonic feedback scenarios. We also explore some of the difficulties in interpreting the small scales in cosmic shear measurements, presenting estimates of the size of several other systematic effects that make inference from small scales difficult, including uncertainty in the modelling of intrinsic alignment on nonlinear scales, `lensing bias', and shape measurement selection effects. For the latter two, we make use of novel image simulations. While future cosmic shear datasets have the statistical power to constrain baryonic feedback scenarios, there are several systematic effects that require improved treatments, in order to make robust conclusions about baryonic feedback.
\end{abstract}





\section{Introduction}\label{sec:intro}
The high galaxy number densities 
of typical weak lensing datasets, and the subsequent large number of galaxy pairs with $\sim$arcminute angular separation, makes shear two-point correlations a powerful probe of the density field on $\lesssim \SI{1}{Mpc}$ physical scales, where density fluctuations are highly nonlinear. The shear two-point signal depends on the matter power spectrum, $\pofk$, which 
describes statistically the two-point clustering of matter as a function of scale (the physical wavevector, $k$) and redshift, $z$. We need to be able to predict $\pofk$ accurately, given a set of cosmological parameters, if we are to infer anything about those cosmological parameters. 

For $k\gtrsim \SI{0.1}{\h\megaparsec^{-1}}$, N-body simulations are required to predict the nonlinear matter clustering
. 
Epic computational demands come from the requirement that the simulations are large enough to include the effects of large-scale power and subdue sampling variance, and have sufficiently high resolution to reach the large $k$ required to make predictions of e.g. the small scale cosmic shear signal (see e.g. \citealt{heitmann10} for discussion of the simulation requirements for matter power spectrum prediction).  To make predictions for a range of different cosmological models, we require the simulations to be re-run many times i.e. a suite of simulations is required. The most advanced example of this sort of suite is the Extended Coyote Universe simulations \citep{heitmann14}, which was used to build a matter power spectrum emulator accurate to $5\%$ up to $k=10h \text{Mpc}^{-1}$ and $z=4$. These types of simulations are often called `dark-matter-only' simulations, although `gravity-only' would perhaps be more appropriate since they do have $\Omega_b>0$, but do not include the effects of non-gravitational physics. As we discuss below, non-gravitational or `baryonic' physics may have a significant effect on the matter clustering on nonlinear scales.

\citet{white04}, \citet{zhan04} and \citet{huterer05} first identified the potential of baryonic physics to contaminate the cosmic shear signal, using simple theoretical models to predict several percent changes in the shear power spectrum at multipoles $l\gtrsim 1000$. \citet{jing06, rudd08, hearin09, guillet10, casarini12} used \textit{hydrodynamic simulations} to account for the many complex baryonic processes such as active galactic nuclei (AGN) feedback, gas cooling and supernovae feedback which affect the matter power spectrum. Hydrodynamic simulations incorporate gas physics by including fluid dynamics as well as gravity, and are consequently more computationally expensive than gravity-only simulations. To fully simulate the relevant baryonic physical processes would require far higher resolution than can currently be achieved for the large volumes required for cosmology, so they are added using `sub-grid' prescriptions. Since we have incomplete understanding of these physical processes, these sub-grid prescriptions need to be calibrated against observables. For example in the state-of-the-art EAGLE simulations \citep{schaye15,crain15}, stellar and AGN feedback efficiency is calibrated to reproduce the observed $z\sim0$ galaxy stellar mass function (GSMF). While this guarantees that the feedback implementation is accurate in its effect on the $z\sim0$ GSMF, it does not guarantee the feedback implementation is accurate in its effect on e.g. the $z\sim1$ GSMF or the nonlinear matter power spectrum. One might conclude that although hydrodynamic simulations can give us indications of the size and scale-dependence of baryonic effects on the matter power spectrum, they are not yet sufficiently advanced to make predictions at the level of accuracy required for precision cosmology.

Various works have made use of the Overwhelmingly Large Simulations (OWLS, \citealt{schaye10}), a suite of hydrodynamic simulations incorporating a variety of baryonic physics scenarios, for assessing the possible impact of baryonic physics on cosmic shear. \citet{vandalen11} measure matter power spectra from the different OWLS simulations which \citet{sembolini11} propagate to the shear two-point functions, finding deviations from the dark-matter-only case as large as $10-20\%$ for shear correlation functions $\xip(\theta=1')$ and $\xim(\theta=10')$.

Most previous cosmic shear studies have either ignored baryonic effects or discarded small scales from their analysis to reduce any potential bias from baryonic effects (see e.g. \citealt{kitching14,maccrann15} for the latter approach).
Recently however, \citet{joudaki16} performed a tomographic analysis of the CFHTLenS \citep{heymans12} data, and marginalised over the possible baryonic feedback on the matter power spectrum, using a one-free-parameter version of the \citet{mead15} halo model (see \sect{sec:baryon_mod} for further details). Unlike this work, their aim is to investigate the much discussed (e.g. \citealt{battye2014,maccrann15,planckcosmo13}) tension with the \planck\ CMB constraints, rather than attempting to differentiate baryonic feedback scenarios, and they do not report constraints on baryonic feedback models. 

\citet{kitching16} also investigate the tension between CFHTLenS and \planck\ by fixing the cosmological parameters to best-fit values from \citet{planckcosmo15}, and constraining various weak lensing nuisance parameters using the CFHTLenS data, including those sensitive to baryonic effects and intrinsic alignments. When allowing a free intrinsic alignment amplitude, they demonstrate a weak preference for a decrement in the matter power spectrum at small scales (compared to the no-baryonic feedback prediction), but no significant evidence for baryonic feedback. 
\citet{harnois14} also use the CFHTLenS data to investigate baryonic feedback by fixing the cosmological parameters to best-fit WMAP9 \citep{wmap9} values, and constraining a 15 free parameter fitting formula describing deviations in the matter power spectrum due to baryonic feedback.

Most recently, \citet{hildebrandt16} use the same prescription as \citet{joudaki16} to marginalise over uncertainty due to baryonic feedback in their cosmic shear analysis of KiDS\footnote{\url{http://kids.strw.leidenuniv.nl}} survey data.
\citet{viola15} also use KiDS weak lensing data, but use the tangential shear signal around galaxy groups. They compare the group mass as a function of BCG luminosity to predictions from the OWLS simulations, and observe a decrement in group mass at high luminosity that favours the prediction of the OWLS simulation containing AGN feedback.

\begin{figure}
\includegraphics[width=\columnwidth]{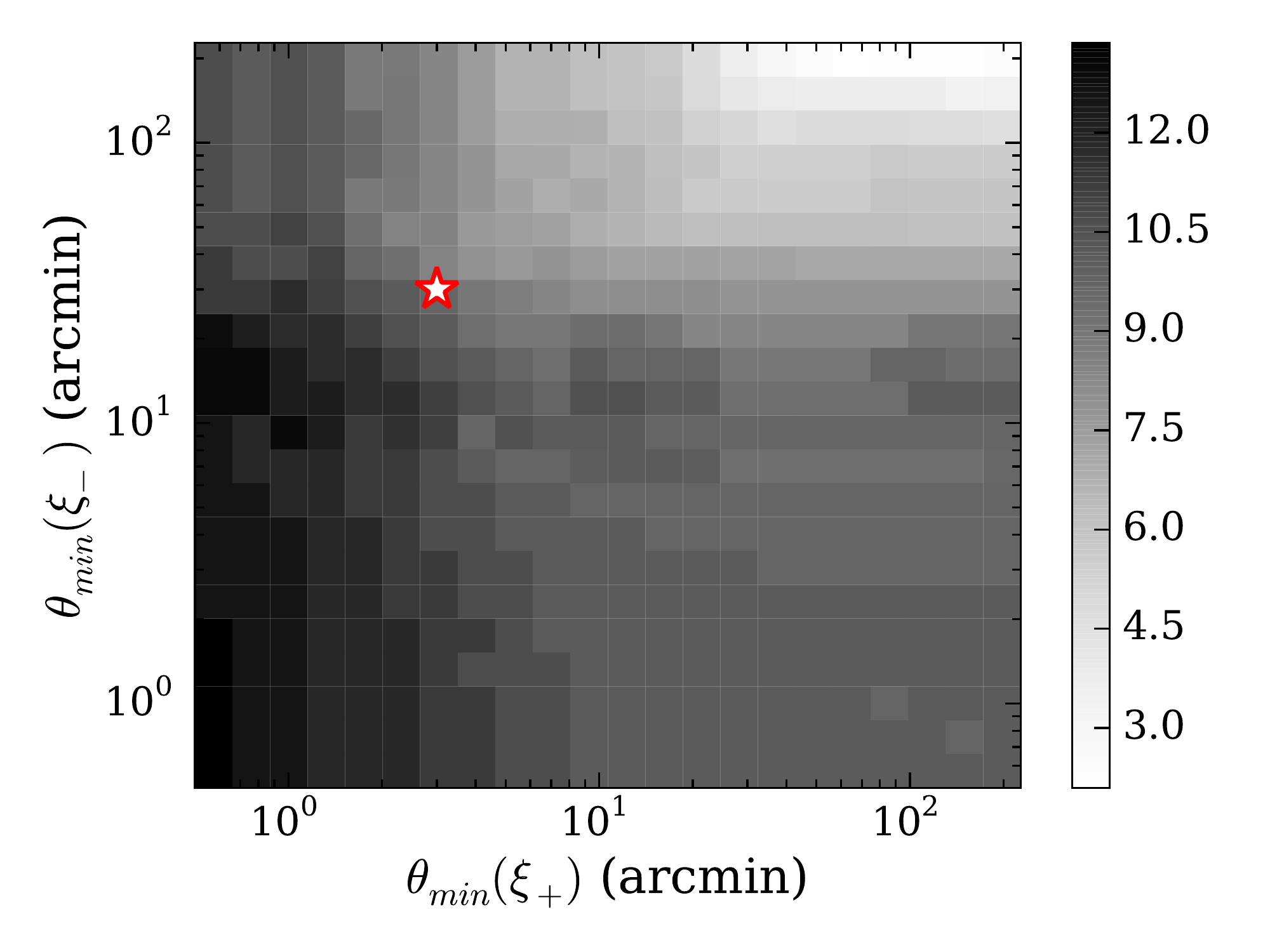}
\caption[DES-SV cosmic shear SNR as a function of minimum scale.]{\snr of the DES-SV non-tomographic correlation functions $\xipm(\theta)$, as a function of the minimum scale use in $\xipm$, $\theta_{\rm{min}}(\xipm)$. The red outlined star marks the minimum scales used in \citetalias{DES15}; clearly there is further signal to be exploited by reducing the minimum scales used.}
\label{fig:small_scale_snr}
\end{figure}

\citet{DES15} (\citetalias{DES15} henceforth) presented cosmological constraints from \SI{150}{\degsq} of Dark Energy Survey Science Verification (DES-SV) data. Using DECam \citep{decam}, the final DES survey will image an area around thirty times this size.
 The DES-SV galaxy shear catalogues are described in \citet{jarvis15}, the photometric redshift estimates in \citet{bonnett15}, and the shear two-point measurements in \citet{Be15}.
They used the matter power spectra from \citet{vandalen11} to calculate a set of minimum angular scales on which to use the measured shear correlation functions, that would reduce any bias due to baryons to below the level of the statistical errors.
The present paper is motivated by the significant signal-to-noise (\snr) that this procedure wastes.  \fig{fig:small_scale_snr} demonstrates this; it shows the total \snr\ of the DES-SV non-tomographic shear correlation functions $\xipm(\theta)$, as a function of $\theta_{\mathrm{min}}(\xipm)$, the minimum scales used in $\xipm(\theta)$. The red star marks the minimum scales used in \citetalias{DES15}, and it's clear that more \snr\ (from $\sim8$ up to $\sim13$) can be gained by reducing these minimum scales. 
Even if astrophysical uncertainties are such that we cannot reliably infer cosmological parameters from the small scale cosmic shear signal, it may be possible to learn about the astrophysical effects themselves.
Therefore it is tempting to try and exploit the extra \snr\ by including the small scales, and attempting to model the effects of baryons.


In \sect{sec:meas} we present new DES-SV cosmic shear measurements, using the galaxy shape catalogues described in \citet{jarvis15}. These measurements are extended to smaller scales than those used in \citet{Be15} and \citetalias{DES15}. In \sect{sec:baryon_mod} we review some methods for modelling or parametrising the effect of baryons on the matter power spectrum, including the extended halo model of \citet{mead15}, and apply the \citet{mead15} model to these new cosmic shear measurements. We also forecast the potential of the final DES 5-year (Y5) data to constrain this model.

Although baryonic effects may be the largest, there are several additional systematic effects that arise on small scales, which we describe in \sect{sec:other_sys}. We estimate several of these, and test their impact on the DES Y5 forecasted results. Firstly, the observed (two-point) cosmic shear signal is usually considered to be sensitive only to second order correlations in the underlying density field (and hence can be written as an integral over the matter power spectrum, see e.g. \citet{bartelmann01}). In \sect{sec:red_mag}, we describe the corrections at third order in the density field that become significant on small scales.
Meanwhile, the removal of blended objects during shape measurement can introduce a selection bias on the cosmic shear signal at small scales \citep{hartlap11}; we call this `blend-exclusion bias', and investigate this effect using image simulations in \sect{sec:ufig}. A further possible complication in interpreting the small-scale signal is intrinsic alignments, for which the successful large-scale models such as the (nonlinear-)linear alignment model \citep{catelan01, HS04, BK07} are likely to break down; we discuss this in \sect{sec:IAs}.
Finally, we note that constraints from cosmic shear will of course be cosmology dependent, and one would expect the constraints on baryonic physics to be most degenerate with other phenomena that produce a scale-dependent change in the matter power spectrum, for example massive neutrinos. We investigate this degeneracy in \sect{sec:cosmo}.

\section{Small-scale extended DES SV shear correlation functions}\label{sec:meas}

\begin{figure*}
\includegraphics[width=\linewidth, trim=-1cm 0cm 0cm 3cm, clip=true]{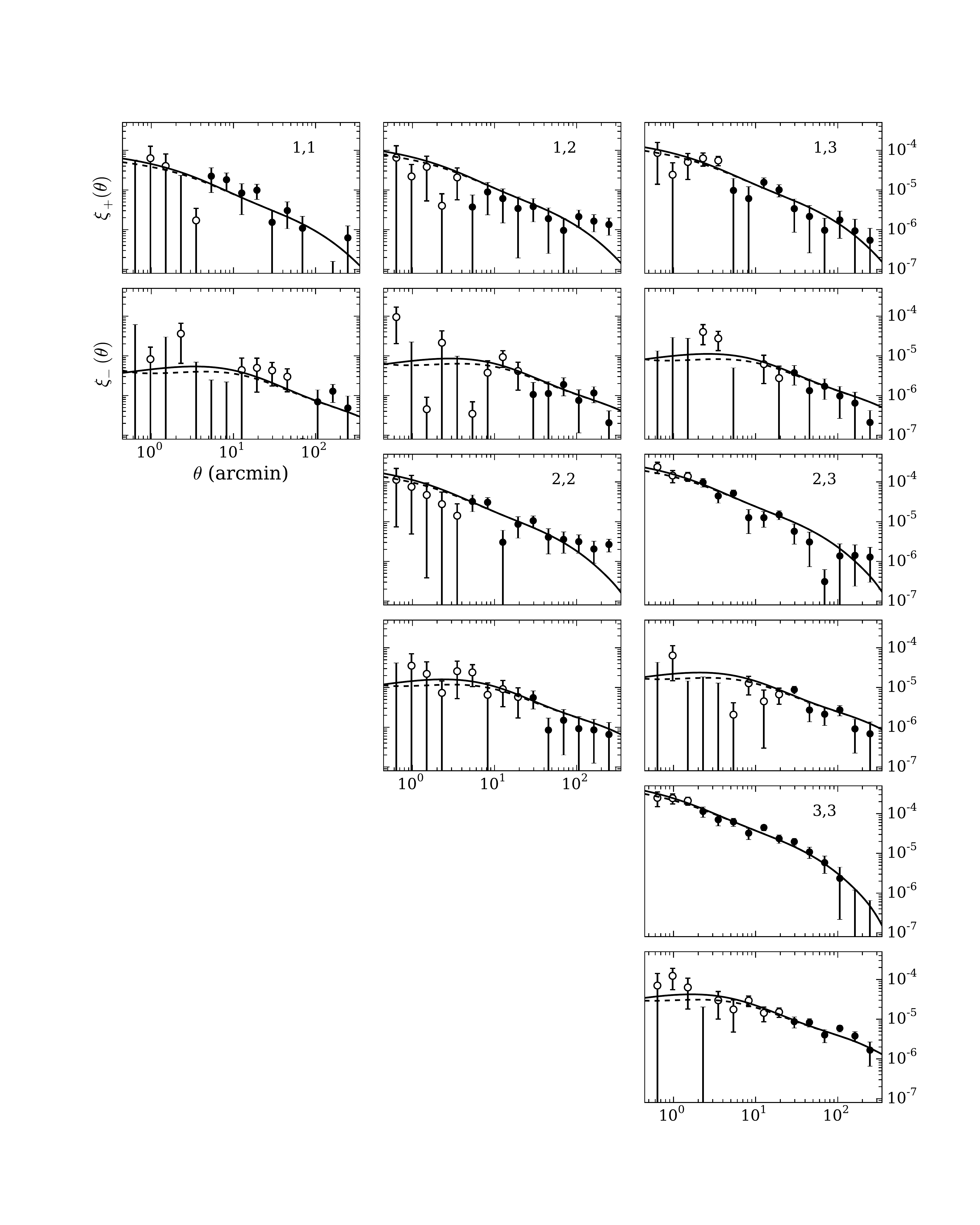}
\caption[Extended DES-SV correlation functions.]{Shear correlation functions, $\xipm$ from DES-SV data, now using a data vector extended to smaller scales than in \citetalias{DES15} (open symbols indicate these smaller scales). The redshift bin pairing is shown in the upper right corner of each $\xip$ panel, and the corresponding $\xim$ measurement in the panel below. The solid line in each panel is the prediction using the \planck\ 2015 cosmology described in \sect{sec:SV}, and using the \citet{takahashi2012} version of \halofit\ \citep{smith03}. The dashed line shows the prediction from the OWLS AGN matter power spectrum (see \sect{sec:meas} for details.)}
\label{fig:xi_SV_ext}
\end{figure*}

In this section we extend the DES-SV shear correlation function measurements to smaller scales.
Figure \ref{fig:xi_SV_ext} shows measurements of the shear correlation functions $\xipm$ in 15 angular bins between 0.5 and 300 arcminutes, in the same three redshift bins described in \citet{Be15} and \citetalias{DES15}. We follow \citetalias{DES15} by excluding angular scales greater than 60 arcminutes from $\xip$, to reduce the impact of additive systematics.
There is a significant signal at scales down to 0.5 arcminutes, particularly for the highest redshift bin. At scales less than a few arcminutes shape-noise, which arises from the uncorrelated intrinsic (unsheared) shapes of galaxies, is the dominant contribution to the covariance, so the data points are only weakly correlated. We conservatively choose 0.5 arcminutes as the smallest separation used. While there still may be some signal below this, shape measurement systematics due to blending may become important. 

The original \citetalias{DES15} cosmic shear analysis used
a covariance matrix calculated from 126 mock survey simulations, as described in \citet{Be15}. \citet{Be15} discussed the limitations on the accuracy of the parameter constraints that can be achieved when the number of simulation realisations is not much greater than the number of data points in the data vector \citep{taylor2013,DodelsonSchneider13}. For the extended tomographic data vector that we use in this work, this requirement is clearly not satisfied. 
We therefore use a covariance inferred from lognormal realisations of the lensing convergence across the survey area.

On large scales, the weak lensing convergence field (and therefore shear fields) is well described by Gaussian statistics, so a simple approximation to the cosmic shear covariance can be obtained by generating many Gaussian random shear fields with the expected shear power spectrum, and computing a sample covariance matrix using the same method as on the mocks. Since generation of the Gaussian realisations is very fast, the covariance uncertainty due to having a finite number of realisations can be made negligible. On smaller scales, the convergence field is sensitive to nonlinearities in the density field, and the Gaussian approximation is no longer a good approximation. However, \citet{taruya02} and \citet{takahashi11} demonstrate that lognormal statistics provide a good description of the convergence field, while \citet{hilbert11} demonstrate that a covariance matrix obtained under the lognormal approximation results in very accurate confidence intervals on cosmological parameters, even when using sub-arcminute scales. \citet{clerkin16} found that the probability distribution function of both galaxy overdensity and convergence in the DES-SV data could be well approximated as lognormal, although they only investigated large ($>$10 arcminutes) scales.

It is probable that the non-Gaussian terms in the covariance will be more accurately accounted for using the halo model \cite{peacock00,seljak00}, as in e.g. \citet{sato09,takada13,cosmolike}, which is a more physically motivated analytic description for the non-Gaussianities. However, accounting for the survey mask is likely to be more difficult in this approach. 


\section{Modelling baryonic effects on the matter power spectrum}\label{sec:baryon_mod}

\subsection{Modelling approaches}

We know from hydrodynamic simulations that baryonic physics can have a significant effect on the matter power spectrum at small scales. However, as described in \sect{sec:intro}, given the uncertainty in what physical processes to add to the simulations at the sub-grid level (as well as the uncertainties due to different implementations of the same sub-grid physics), the magnitude, scale-dependence and redshift dependence of the effect is very uncertain. In order to extract any information from the small scales, a model or nuisance parameterisation that is sufficiently flexible to describe the baryonic effects, is required. Judging how flexible is `sufficiently' flexible is always a challenge when assessing the suitability of a nuisance parametrisation. The fact that a nuisance parameterisation is required means that we lack knowledge about the physical process. However, deciding on a parameterisation and priors on the nuisance parameters requires assumptions (presumably based on some knowledge) about the same physical process.

In the case of baryonic effects on the matter power spectrum, hydrodynamic simulations arguably provide a level of knowledge sufficient to provide the basis of a nuisance parameterisation, or usefully test the flexibility of a modelling approach. The proposal of \citet{eifler14} makes this assumption; they propose using principal component analysis (PCA) to identify modes with the most variance between multiple simulations with different baryonic treatments. These modes can then be projected out of the analysis, providing a way of retaining only the information unaffected by baryonic effects that is more sophisticated than e.g. simply imposing a minimum angular scale.

More recently, \citet{foreman16} present a method for using cosmic shear to constrain the matter power spectrum in a fairly model independent way; by allowing deviations from the dark-matter-only $\pofk$ at grid points in $k$ and $z$. They demonstrate that using PCA to identify the best-constrained modes allows a decrease in the number of free parameters used, while retaining most of the information on any power spectrum deviation.

Another approach is to use a theoretical model for the matter power spectrum, with some physically motivated free parameters to account for possible baryonic effects. \citet{zentner08,hearin09} showed that the effect of baryons on the matter power spectrum could be qualitatively reproduced in the halo model framework. The halo model \citep{seljak00,peacock00} is an analytic model for the matter distribution in the Universe, that, given its simplicity, is extremely successful at reproducing the matter power spectrum, even on nonlinear scales. The model assumes that all matter is contained in spherical halos. The halo radial density profile is assumed to depend only on the mass of the halo. The statistical properties of the matter field are then set by three inputs: (i) the relation between the halo  density profile and mass, (ii) the number density of halos of a given mass, and (iii) the large scale distribution of halos, which just depends on the linear matter power spectrum. The halo density profile is usually taken to be the \textit{NFW} profile \citep{navarro96}, which for a given mass, has one free parameter, the concentration. Input (i) is then the `concentration-mass relation'. Input (ii) is the halo mass function, the fraction of halos in a given mass range. Both the concentration-mass relation and the halo mass function can be calibrated using N-body simulations.

\citet{mead15} use the halo model as a basis for which to tackle the problem of predicting the nonlinear matter power spectrum. They first implement various adjustments to the basic halo model described above which are required to accurately predict the dark matter-only matter power spectrum. With these adjustments, they achieve a 5\% matter power spectrum accuracy for $k\leq 10 h \text{Mpc}^{-1}$, $z \leq 2$, which they judge by comparison with the Coyote Universe simulations. In fact, the accuracy exceeds $2\%$ apart from around scales of $k=\SI{0.2}{\h/\megaparsec}$ where damping of the BAO is important, which they do not attempt to model. 

\begin{figure}
\centering
\includegraphics[width=\columnwidth, trim=0cm 0cm 0cm 0cm, clip=true]{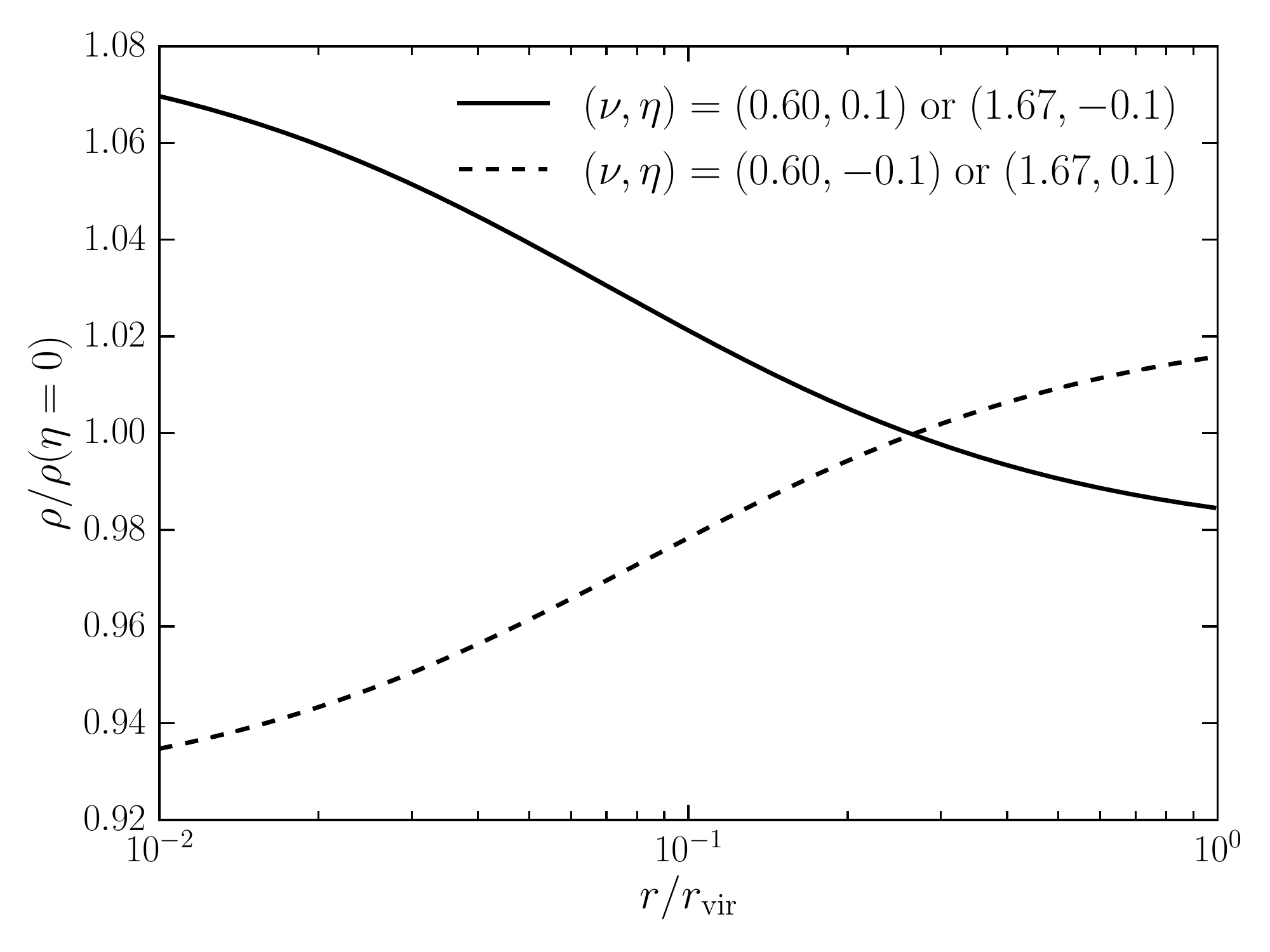}
\caption[]{The fractional change in the halo density profile (as a function of radius in units of the virial radius of the $\eta=0$ halo) due to non-zero $\eta = \eta_0 - 0.3\sigma_8(z)$, for a high mass ($\nu>1$) and low mass halo ($\nu<1$). $\eta_0$ is one of the two free parameters in the \citet{mead15} halo model (see \sect{sec:baryon_mod} for more details). }
\label{fig:eta}
\end{figure}

They further extend this halo model to account for baryonic effects. We will refer to this extended halo model as the `M+15' model. Since baryonic physics are likely to change the internal structure of halos, but have a lesser effect on their positions or total masses, they propose two extra nuisance parameters to allow for the former. Firstly, they allow to vary $A$, the amplitude in the concentration-mass relation i.e. increasing $A$ makes halos of all masses more concentrated. The second free parameter is $\eta_0$, which they call the `halo bloating parameter', since it produces a (mass-dependent) bloating of the halo profile. To describe the effect of $\eta_0$, we first define
\be
\nu \equiv \frac{\delta_c}{\sigma(R(M))},
\ee
where $\delta_c$ is the linear theory overdensity collapse threshold and $\sigma(R(M))$ is the linear theory density variance in spheres of radius $R$ that on average contain mass $M$. So $\nu<1$ halos can be categorized as low mass, while $\nu>1$ halos can be categorized as high mass.
The halo profile in Fourier space, $W(k,M)$ is modified as 
\be
W(k,M) \rightarrow W(\nu^\eta k, M), 
\ee
where $\eta = \eta_0 - 0.3\sigma_8(z)$. The result is that low mass ($\nu<1$) halos are more concentrated when $\eta>0$ and more bloated when $\eta<0$, while conversely high mass ($\nu>1$) halos are more bloated when $\eta>0$ and more concentrated when $\eta<0$. \fig{fig:eta} shows the fractional change in the density profile of low and high mass halos for positive and negative $\eta$. The figure demonstrates that the change in a $\nu=0.6$ (i.e. low mass) halo profile due to setting $\eta=0.1$ is the same as the change in a $\nu=1.67$ (i.e. high mass) halo profile due to setting $\eta=-0.1$.

They test this parameterisation by fitting the model to matter power spectra from three of the OWLS simulations, and in all cases achieve similar accuracy ($\sim< 2\%$ up to $k = \SI{10}{\h/\megaparsec}$, apart from the BAO wiggles) in the matter power spectrum as for the dark-matter-only case (at the cost of these two extra nuisance parameters). The three OWLS simulations used are the `REF', `DBLIM' and `AGN' simulations. The `REF' simulation contains radiative cooling and heating, stellar evolution, chemical enrichment, stellar winds and supernova feedback. The `AGN' simulation is similar to the `REF' simulation but additionally contains feedback from AGN. The `DBLIM' simulation is again similar to `REF' but has additional supernovae energy in wind velocity, and a top heavy initial mass function at high pressure. See \citet{schaye10} for detailed description of these simulations. 
We note here that the recent study of \citet{cui16} indicates that the methodology used in the OWLS simulations suite should be thought of as one of several that have not yet demonstrated convergence. That study shows that the even the \emph{sign} of the effect on halo internal mass structure varies among simulation methods. Our use of OWLS as a reference in this work should be considered as illustrative of the potential magnitude of these complex effects.

To implement the M+15 model, we use \textsc{HMCode}\footnote{\url{https://github.com/alexander-mead/HMcode}}, code made publicly available by \citet{mead15}, and included in the \cosmosis\ \citep{cosmosis} package. We use the \cosmosis\ framework for all parameter inference in this work.

\subsection{Constraints from DES-SV}\label{sec:SV}

\begin{figure}
\centering
\includegraphics[width=\columnwidth, trim=0cm 0cm 0cm 0cm, clip=true]{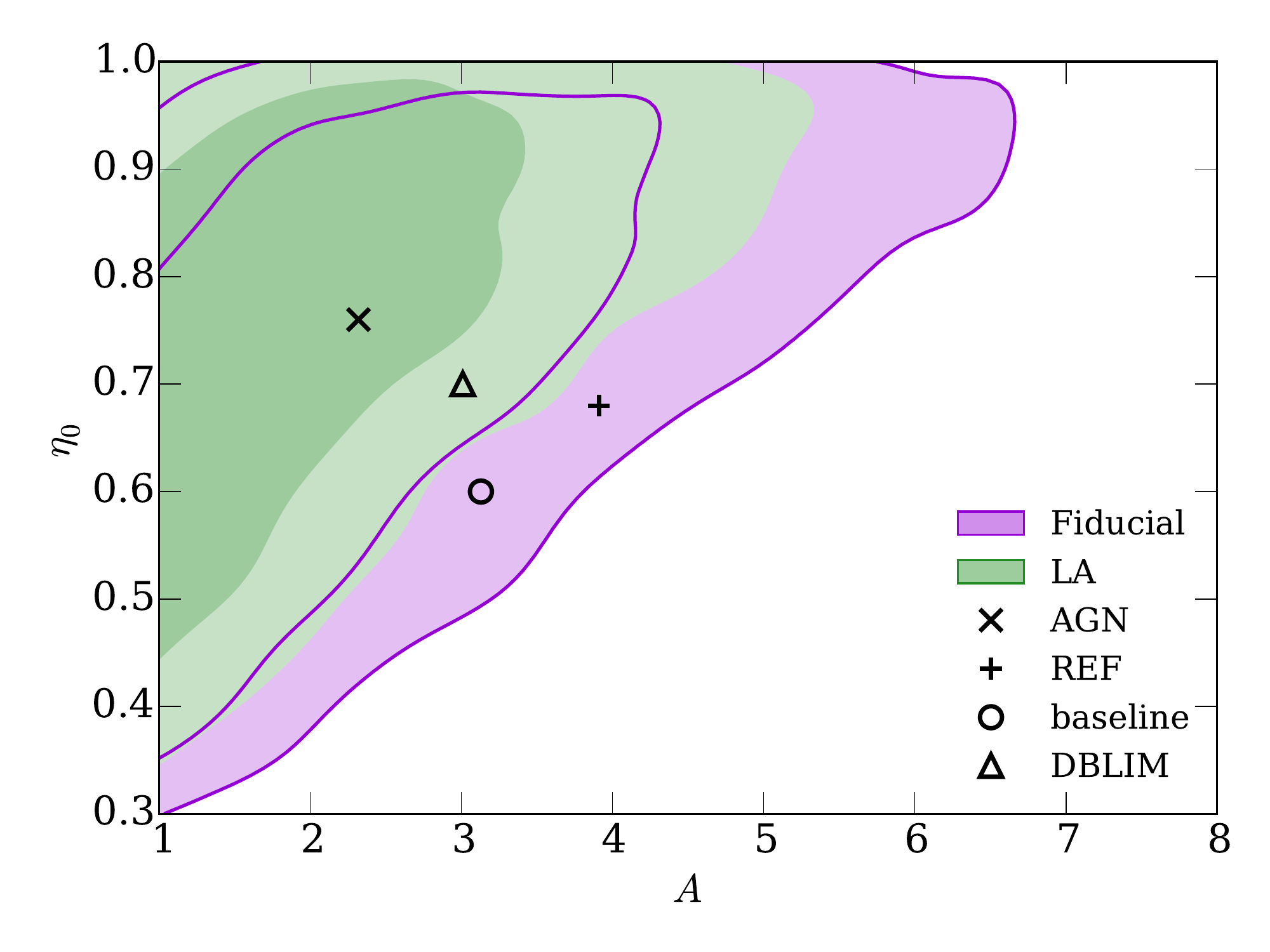}
\caption[DES-SV constraints on the M+15 halo model.]{DES-SV cosmic shear one and two-sigma constraints on the two nuisance parameters of the M+15  halo model \citep{mead15}. The data vector has been extended to smaller scales than the original analysis \citepalias{DES15}. 
The purple filled and outlined contour is the fiducial analysis, the same three redshift bins as \citetalias{DES15}, and angular scales in the range 0.5 to 300 arcminutes. The green filled contour models galaxy intrinsic alignments using the linear alignment model \citep{catelan01,HS04} rather than the nonlinear linear alignment model \citep{BK07} used for the fiducial analysis.
The black markers show the best-fit parameters for several different OWLS simulations, calculated by \cite{mead15}. The plus is `REF', the cross is `AGN', the triangle is `DBLIM' and the circle is for no baryonic effects (see \citealt{schaye10,mead15} for descriptions of the OWLS simulation names).}
\label{fig:SV}
\end{figure}

\fig{fig:SV} shows the constraints on $A$ and $\eta_0$ from the DES-SV cosmic shear measurements described in \sect{sec:meas}. As well as the two halo model parameters, the same set of systematics parameters as used in \citetalias{DES15} are marginalised over: a redshift bin shift parameter per redshift bin, $\delta z_i$; a multiplicative shear bias per redshift bin, $m_i$; and an intrinsic alignment amplitude, $A_{\rm{IA}}$. For the purple contour labelled `fiducial', the intrinsic alignment model used is the `nonlinear-linear alignment' (NLA) model of \citet{BK07}, which was the fiducial model used in \citetalias{DES15}. As in \citetalias{DES15}, Gaussian priors of width 0.05 are used for the $\delta z_i$ and $m_i$, and a uniform prior [-5,5] on $A_{\rm{IA}}$ is used. Cosmological parameters are fixed to the \citet{planckcosmo15} `\planck\ TT + lowP' values. The allowed ranges of the parameters $A$ and $\eta_0$ are those plotted, which \citet{mead15} showed to be comfortably wide enough to span the space of simulations considered there. 

Although the constraints from DES SV are fairly weak, the high $A$, low $\eta_0$ region of the parameter space is strongly disfavoured. Shown as black marks are the best-fit halo model parameters to various cosmological simulations, as estimated by \citet{mead15}: The circle is the $(A,\eta_0)$ which they find to be the best-fit to the Coyote Universe simulations, which do not contain baryonic feedback effects; we call this the `baseline' case. The plus is the best-fit $(A,\eta_0)$ for the OWLS `REF' simulation, which contains radiative cooling and heating, stellar evolution, chemical enrichment, stellar winds and supernova feedback (see \citet{schaye10} for detailed descriptions of the OWLS simulations). The cross is the best-fit $(A,\eta_0)$ for the OWLS `AGN' simulation, which is similar to the `REF' simulation, but additionally contains feedback from AGN. The triangle is the best-fit $(A,\eta_0)$ for the OWLS `DBLIM' simulation, which is again similar to `REF', but has additional supernovae energy in wind velocity, and a top heavy initial mass function at high pressure.

Note that we do not constrain the likelihood of the OWLS simulations directly - rather the parameters of the M+15 halo model, which we assume is flexible enough to account for a wide range of baryonic effects. So when we say e.g. ``the AGN model is disfavoured with $X\%$ confidence'', we really mean the $(A,\eta_0)$ preferred by the OWLS AGN simulation is disfavoured with $X\%$ confidence. Given the success of the M+15 model in encapsulating the different OWLS simulations, we believe this is a reasonable way to report constraints, but it is important to be clear that our constraints are on the halo model parameters, rather than on the OWLS simulations directly. 

For the AGN model, the preferred $(A,\eta_0)$ lies on the contour of equal-probability containing $22.8\%$ of the posterior probability. We define the quantity $C_{\rm{M}}$, for baryonic model M with $(A,\eta_0)=(A^M,\eta_0^M)$, as the percentage of the posterior weight contained within the contour of equal posterior on which $(A^M,\eta_0^M)$ lies. So $C_{\rm{AGN}} = 22.8\%$.
A model M with $C_{\rm{M}}$ of $95\%$ would be considered disfavoured with $95\%$ confidence. We find $C_{\rm{baseline}}=82.9\%$, $C_{\rm{DBLIM}}=52.0\%$ and $C_{\rm{REF}}=86.9\%$, so none of the models are strongly disfavoured by the DES-SV cosmic shear data. 

\section{Predicted baryonic constraints from DES cosmic shear and small-scale systematics}\label{sec:other_sys}

\begin{figure*}
\includegraphics[width=0.45\linewidth, trim=0cm 0cm 0cm 0cm, clip=true]{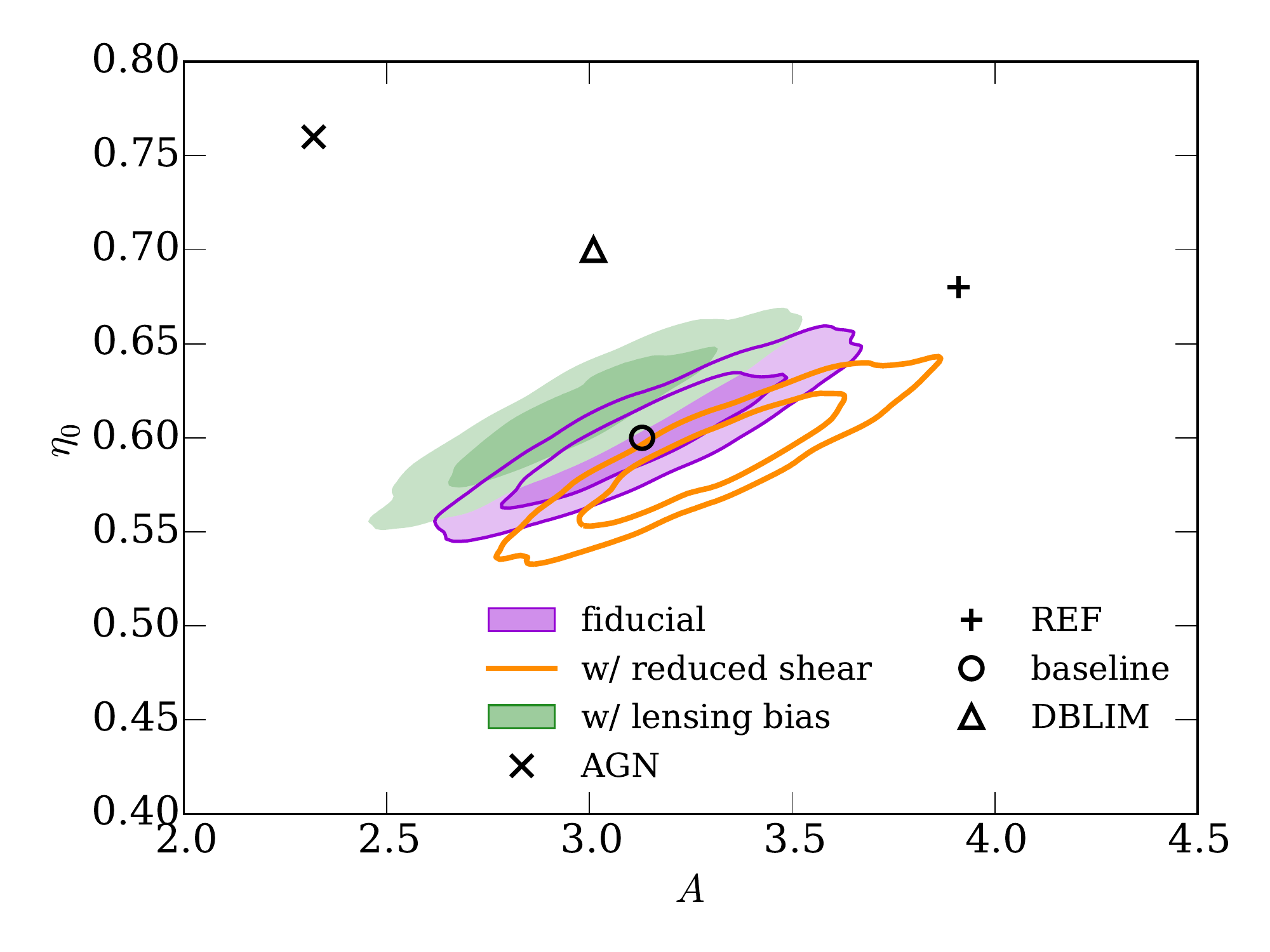}
\includegraphics[width=0.45\linewidth, trim=0cm 0cm 0cm 0cm, clip=true]{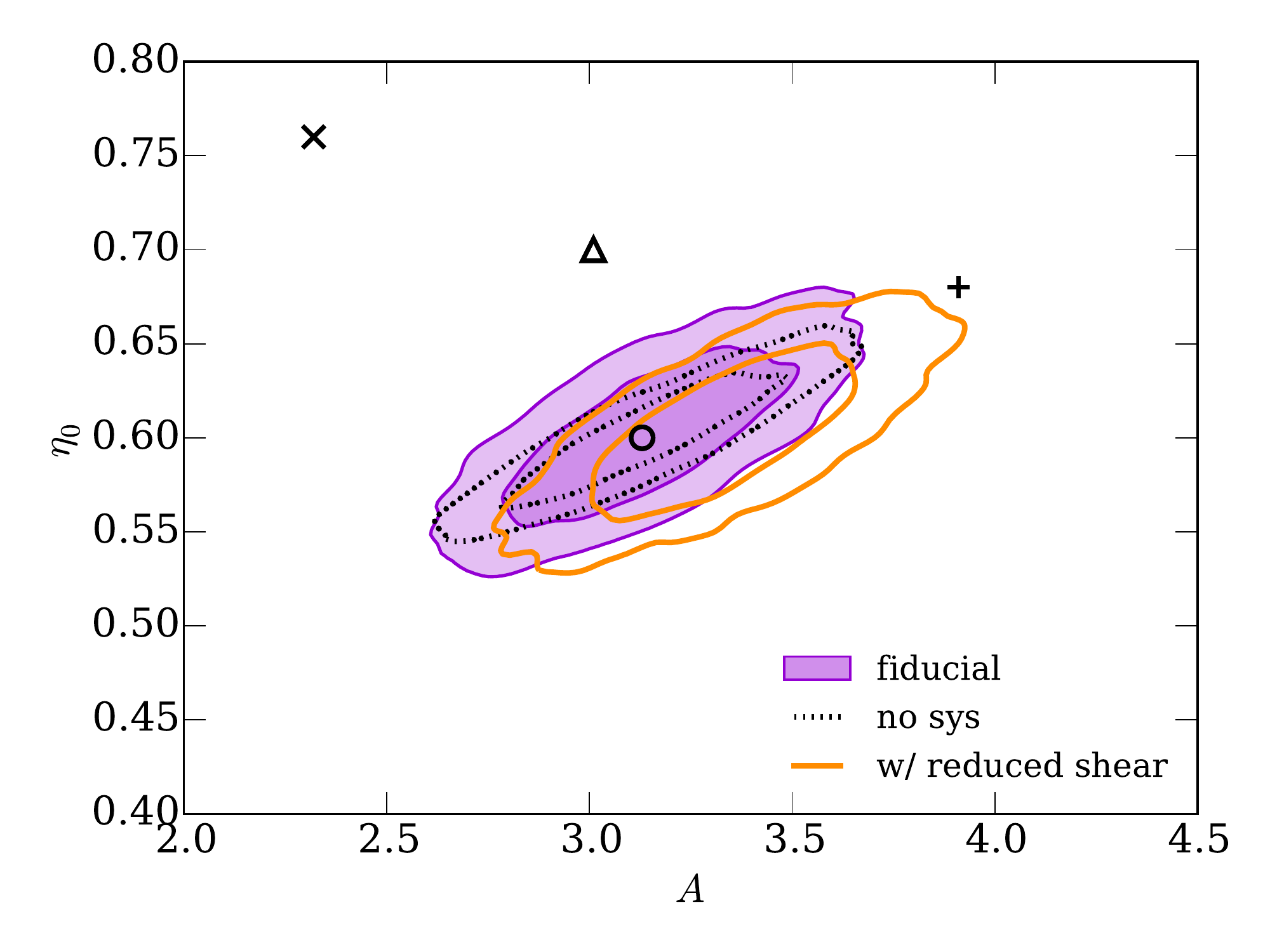}
\caption[DES Year 5 Mead+15 model forecasts.]{
Expected constraints (assuming the baseline model) on the Mead+15 model parameters from DES Year 5 cosmic shear. We assume a tomographic data vector with 5 redshift bins, using an angular range $0.5<\theta<300$ arcminutes. \textit{Left panel}: The weak lensing nuisance parameters described in \sect{sec:SV} are \textit{not} marginalised over. The purple (outlined and filled) contours show the constraints with no systematics added to the simulated data vector, hence the input halo model parameters are correctly recovered.  For the green filled (orange outlined) contours, we include a correction to the simulated data vector for reduced shear (lensing bias). When fitting the simulated data vector, we do not include either correction, hence the contours are shifted, and the inferred halo model parameters are somewhat biased. \textit{Right panel}:  The purple and orange contours are the same as in the left panel, except now marginalising over the 11 weak lensing nuisance parameters (a  $\delta z_i$ and $m_i$ per redshift bin and an intrinsic alignment amplitude $A_{\rm{IA}}$), hence the constraints on the halo model parameters become weaker, and the bias due to ignoring the reduced shear correction becomes less significant. For comparison, the black dotted line shows the constraints without marginalising over the WL nuisance parameters (i.e. the purple contour in the left panel). In both panels, the black markers are the same as in \fig{fig:SV}.}
\label{fig:y5}
\end{figure*}

While current cosmic shear data such as DES-SV only weakly constrains models of baryonic physics and its effects on the Universe's matter distribution, upcoming datasets will have far greater statistical power. The final Dark Energy Survey dataset, which we call `Y5', since it will be composed of five years of data, will be around thirty times larger in area than DES-SV. The purple filled/outlined contour in \fig{fig:y5} shows the expected constraints on the M+15 halo model parameters from Y5 cosmic shear data. To perform this forecast we use as a `simulated' data vector a theoretical prediction, and then run an MCMC parameter estimation analysis, as would be performed on a measured data vector. The simulated data vector has no baryonic physics added.
The covariance and data vector are the same as that used in Foreman et al. 2016, which assumes a 5 tomographic bin analysis, over the angular range $0.5 < \theta < 300'$, with 8 galaxies per square arcminute, and an area of $\SI{5000}{\degsq}$. The covariance matrix was computed using \textsc{CosmoLike} \citep{cosmolike,krause16}. Again (and unless otherwise specified), we fix cosmological parameters to the \citet{planckcosmo15} values described in \sect{sec:SV} (we explore variations in cosmological parameters, including the neutrino mass, in \sect{sec:cosmo}). 

In the left panel, no weak lensing systematics nuisance parameters (i.e. the $\delta z_i$, $m_i$ and $A_{\rm{IA}}$ described in \sect{sec:SV}) are marginalised over. In the right panel these systematics parameters are included, although we now use Gaussian priors of width 0.02 for the $\delta z_i$ and $m_i$, which we hope will be justified by higher quality data and improved data reduction tools. Even without such improvements, it is likely that future DES analyses will combine shear two-point measurements with galaxy-galaxy lensing and galaxy clustering measurements which will tighten constraints on systematic parameters (see e.g. \citealt{joachimi10b,zhang10}).
In order to make robust conclusions about baryonic physics, we must ensure that any other uncertainties or systematic biases in the small-scale cosmic shear signal are accounted for. The green filled contour in \fig{fig:SV} shows an example of this. For these contours an alternative model of galaxy intrinsic alignments is assumed, the linear alignment model (\citet{catelan01,HS04}, see \sect{sec:IAs} for more details). When this intrinsic alignment model is assumed, the REF and baseline M+15 halo model parameters (the `+' in \fig{fig:SV}), are now disfavoured, with $C_{\rm{baseline}}=97.0\%$ and $C_{\rm{REF}}=97.2\%$. This is a simple demonstration that even with DES-SV data, including uncertainties in the intrinsic alignment modelling is important.

In this section we discuss various theoretical/systematic uncertainties that can potentially bias conclusions from small scale cosmic shear measurements, including intrinsic alignments (\sect{sec:IAs}). We use the Y5 forecast to quantify the importance of the various systematic effects. In particular, we calculate the credible interval, $\ctruth$ in the $A-\eta_0$ plane, on which the true ($A,\eta_0$) (i.e. those used to generate the simulated data vector) lie, when we include a particular systematic in the simulated data vector, but do not include it in the modelling. A $C_{\rm{truth}}$ value of 90\% would indicate that ignoring that systematic would result in the true values of ($A,\eta_0$) being ruled out with 90\% confidence. So $C_{\rm{truth}}$ quantifies the severity of the bias caused by a particular systematic effect.

\subsection{Reduced-shear and lensing bias}\label{sec:red_mag}

In this section we consider two contributions to the observed cosmic shear signal that arise from third-order correlations of the convergence or equivalently third order in the gravitational potential, $\Psi$ (usually only the second-order correlations are considered, in which case the cosmic shear signal can be written as a projection of the matter power spectrum, see e.g. \citet{bartelmann01}). \citet{krause10} investigate corrections up to $O(\Psi^4)$, and although the $O(\Psi^4)$ terms will be non-negligible for future surveys, the $O(\Psi^3)$ terms are around an order of magnitude larger, and so we only consider the latter here. The observable in cosmic shear is the two-point correlation of the observed ellipticity, $<\epsilon^{\rm{obs}}\epsilon^{\rm{obs}}>$. It is usually assumed that this is an unbiased estimate of the two-point correlation of the shear $<\gamma\gamma>$. Ignoring intrinsic alignments, we describe below two $O(\Psi^3)$ reasons why this is not quite correct.

Firstly, the ellipticity that we measure is actually an estimate of the \textit{reduced shear}, $g$, which is related to the shear via
\be
g=\frac{\gamma}{1-\kappa} \approx \gamma(1+\kappa).
\ee
This requires a `reduced shear' correction to the predicted signal, which is derived in Appendix \ref{app:shear3}, following \citet{shapiro09}.

Secondly, we only observe the shear at the position of galaxies, so when we compute a `shear' two-point correlation function, 
we are effectively computing the correlation function of the galaxy density-weighted reduced shear, $g^{\rm{obs}}$, given by
\be
g_{\text{obs}} = (1+\dobs)g
\ee
where $\dobs$ is the observed galaxy overdensity at a particular point in space. 
We consider two ways in which an observed galaxy overdensity at angular coordinate $\vect{\theta}$ and radial coordinate $\chi$ can arise (apart from random fluctuations). Firstly there could be an overdensity in the galaxy number at $(\vect{\theta},\chi)$ e.g. if there is a cluster there. Secondly, there could be a change in the number density of galaxies that we can observe, due to lensing magnification e.g. if there is a cluster at $(\vect{\theta},\chi'<\chi)$. The first leads to the `source-lens clustering' \citep{bernardeau98,hamana02}, which is zero in the Limber limit (see Appendix \ref{app:shear3}). The second effect produces what is known as \textit{lensing bias} (or sometimes `magnification bias'), and we derive an expression for it in Appendix \ref{app:shear3}, following \citet{schmidt09}.

\subsubsection{The reduced-shear correction}
\begin{figure}
\centering
\includegraphics[width=0.95\columnwidth, trim=0cm 0cm 0cm 0cm, clip=true]{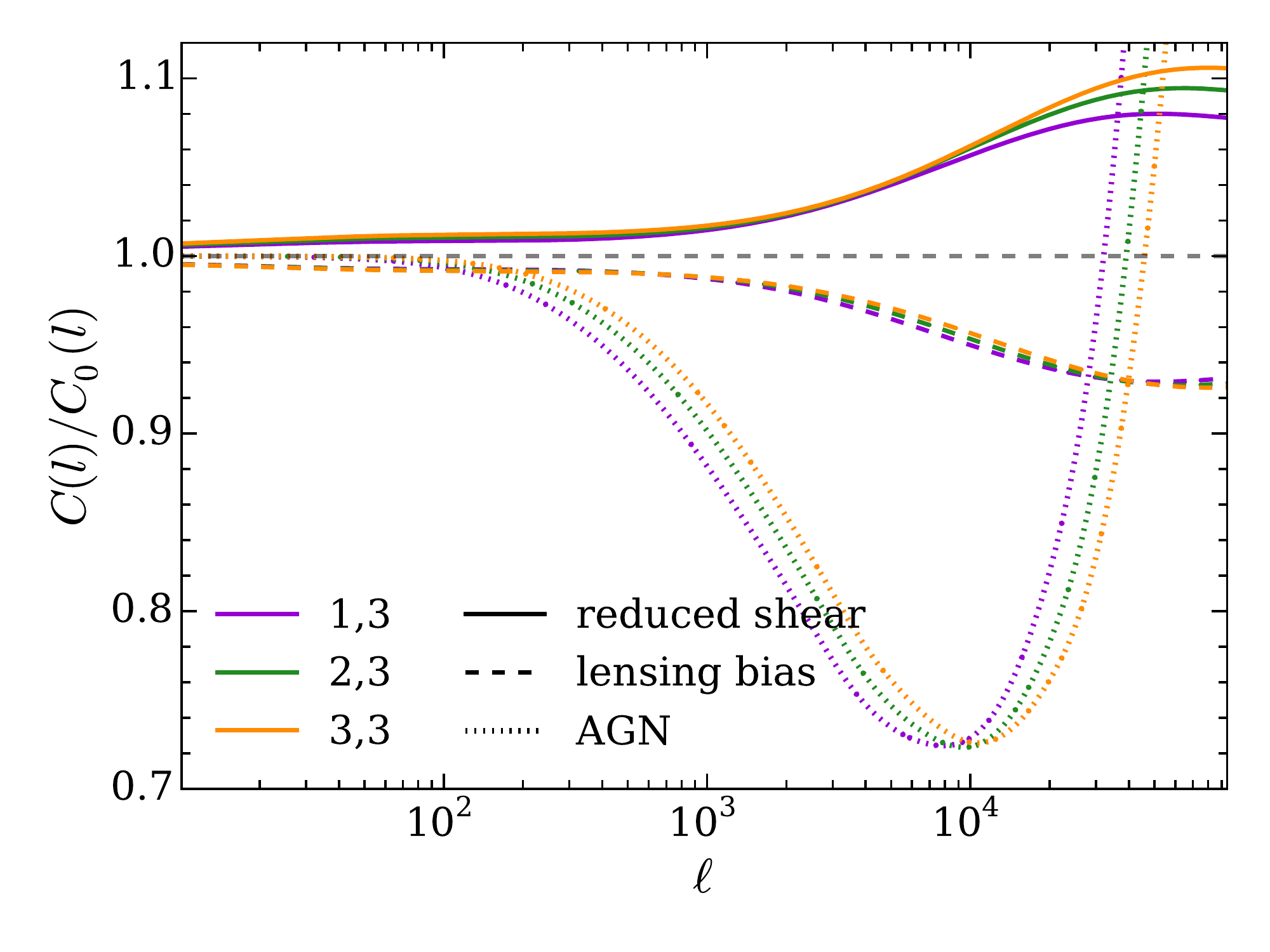}
\caption[Reduced-shear and lensing-bias contributions to the shear power spectra.]{The fractional difference in the shear power spectrum due to reduced-shear (solid lines) and lensing-bias (dashed lines) are compared to that from the OWLS AGN model (dotted lines). We used the DES-SV tomographic redshift distributions, and for clarity only show the correlations with the highest redshift bin.} 
\label{fig:red_mag_Cl}
\end{figure}

From \citet{shapiro09}, the reduced-shear correction to the projected shear power spectrum for tomographic bin pairs $i$ and $j$ is given by
\be
\delta_{\rm{red}} C_{ij}^{\kappa}(l)=\
2\int \frac{\dsx{l'}}{(2\pi)^2} \textrm{cos}(2\phi_{l'}-2\phi_{l}) \
B_{ij}(\vec{l'},\vec{l}-\vec{l'},\vec{-l}),
\label{eq:cl_red1}
\ee
where
\begin{multline}
B_{ij}(\vec{l_1},\vec{l_2},\vec{l_3})=\
\frac{1}{2}\int \frac{\dx{\chi}}{\chi^4}W_i(\chi)W_j(\chi)[W_i(\chi)+W_j(\chi)] \\ B_{\delta}(\vec{k_1}, \vec{k_2}, \vec{k_3}; \chi),
\label{eq:cl_red2}
\end{multline}
$B_{\delta}(\vec{k_1}, \vec{k_2}, \vec{k_3}; \chi)$ is the matter bispectrum, and $W_i(\chi)$ is the radial lensing kernel for redshift bin $i$. We use the fitting formula for the matter bispectrum from \citet{scoccimarro01}. \fig{fig:red_mag_Cl} shows the effect of reduced shear on the shear power spectrum, for the same redshift bins as used in the DES-SV analysis in \sect{sec:SV}. The fractional bias is $\sim1\%$ at $l$ of a few hundred, and $\sim10\%$ at $l$ of $10^4$.

\begin{figure*}
\includegraphics[width=\linewidth, trim=0cm 0cm 0cm 0cm, clip=true]{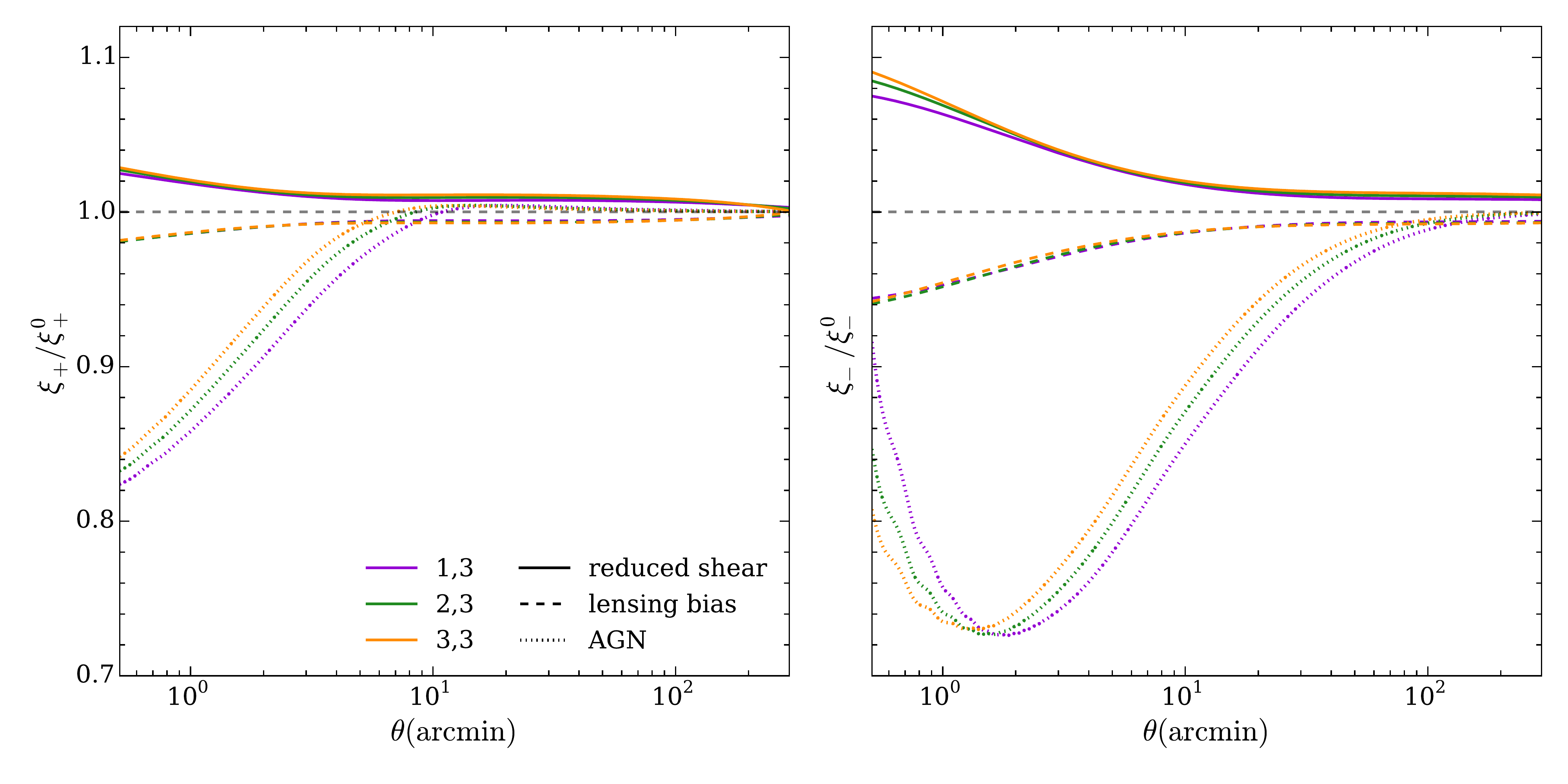}
\caption[Reduced-shear and lensing-bias contributions to the shear correlation functions.]{The fractional difference in the projected shear correlation functions $\xipm$ due to reduced-shear (solid lines) and lensing-bias (dashed lines) are compared to that from the OWLS AGN model (dotted lines). We used the DES-SV tomographic redshift distributions, and for clarity only show the correlations with the highest redshift bin.} 
\label{fig:red_mag_xi}
\end{figure*}

Figure \ref{fig:red_mag_xi} shows the effect of the reduced shear on the shear correlation functions $\xipm$, which at 1 arcminute is $\sim2\%$ for $\xip$ and $\sim8\%$ for $\xim$. Hence the reduced shear correction, although not as large as the effect of baryons in the OWLS AGN model, is non-negligible for small-scale cosmic shear measurements.

\subsubsection{The lensing-bias}\label{sec:lensing_bias}

In Appendix \ref{app:shear3} we derive (drawing heavily on \citealt{schmidt09}) the lensing-bias correction to the shear power spectrum, which for a pair of redshift bins $i$ and $j$ is
\be
\delta_{\text{lensing}}C_{ij}^{\kappa}(l) = \int \frac{\dsx{l'}}{(2\pi)^2} \cos(2\phi_{l'}-2\phi_l) B^q_{ij}(\vec{l'},\vec{l}-\vec{l'},-\vec{l})\label{eq:lens_bias}
\ee
where
\begin{multline}
B^q_{ij}(\vec{l_1},\vec{l_2},\vec{l_3})=\
\frac{1}{2}\int \frac{\dx{\chi}}{\chi^4}W_i(\chi)W_j(\chi)[q_iW_i(\chi)+q_jW_j(\chi)] \\ B_{\delta}(\vec{l_1}/\chi, \vec{l_2}/\chi, \vec{l_3}/\chi; \chi).
\end{multline}
The quantity $q_i$ is given by
\be
q_i=2\beta_f+\beta_r-2\label{eq:q}
\ee
where
\begin{align}
\beta_f \equiv \int \dx{r} \int \dx{f} \frac{\partial\epsilon(f,r)}{\partial(\ln(f))}\Phi(f,r)\\
\beta_r \equiv \int \dx{r} \int \dx{f} \frac{\partial\epsilon(f,r)}{\partial(\ln(r))}\Phi(f,r).
\label{eq:beta}
\end{align}
$\epsilon(f,r)$ is the survey selection function in galaxy flux, $f$, and size, $r$; $\Phi(f,r)$ is the true galaxy distribution in flux and size. We make use of the Balrog simulations \citep{suchyta16} to estimate $\epsilon(f,r)$ and $\Phi(f,r)$. Balrog is a method for simulating observed galaxy catalogues, by injecting simulated objects with known properties into real survey images. The resulting `simulated' images therefore contain many of the important properties of the real data, including observational systematics that would be otherwise difficult to simulate, as well as a small\footnote{Sufficiently small that we need not consider any interaction between the injected objects.} number of extra injected objects. The properties (both morphology and multi-band fluxes) of the inserted objects are based on COSMOS observations, which also have accurate redshifts. By running the same catalogue creation software (in this case \sex, \citet{BertinSExtractor1996}) on these simulated images as is run on the real data, and then repeating the injection and catalog creation process many times over, we can estimate the mapping from the true properties of a galaxy to the properties estimated by \sex\ in our galaxy catalogues. For example, we can estimate the probability of detecting a galaxy with a particular true flux and size, or more generally, the survey selection function as defined above, $\epsilon(f,r)$. We estimate $\Phi(f,r)$ and $\epsilon(f,r)$ as follows.
\begin{enumerate}
\item{We start with the SV \ngmix\ \citep{Sheldon2014} shape catalogue (that was used in \citetalias{DES15}), and the Balrog catalogue used in \citet{suchyta16}, the latter of which contains both `observed' fluxes and sizes i.e. those estimated by \sex, as well as true fluxes and sizes (those used when drawing the simulated objects into the DES images) and redshifts. Note that the observed sizes are PSF-convolved.}
\item{For a given redshift bin of the SV \ngmix\ data, we re-weight the Balrog data to have the same redshift distribution. Then we compute $\Phi(f,r)$ using weighted kernel-density-estimation (KDE) in the true flux and size of these weighted Balrog objects. 
}
\item{We then re-weight the Balrog data to have the same observed flux and size distribution as the \ngmix\ shape catalog, for that particular redshift bin. For the observed flux and size we use the $i-$band \sex~ quantities MAG\_AUTO and FLUX\_RADIUS to do this and make use of the \textsc{hep\_ml} package \footnote{\url{https://github.com/arogozhnikov/hep\_ml}} to perform Gradient Boosting\footnote{\url{https://en.wikipedia.org/wiki/Gradient\_boosting}} re-weighting. With this new set of weights, we again use weighted KDE to estimate $\Phi^{\rm{obs}}(f,r)$, given by
\be
\Phi^{\rm{obs}}(f,r)=\epsilon(f,r)\Phi(f,r).
\ee
}
\item{We then estimate $\epsilon(f,r)$ as $\Phi^{\rm{obs}}(f,r)/\Phi(f,r)$. 
}
\end{enumerate}


\subsubsection{Impact on signal and baryonic constraints}

Having estimated $\Phi(f,r)$ and $\epsilon(f,r)$, the expressions in \eqn{eq:beta} can be calculated and substituted into \ref{eq:q}. We find $q_1 = -1.02 \pm 0.02$, $q_2=-0.79 \pm 0.01$, $q_3=-0.64 \pm 0.01$ (the errors are derived by jackknifing the galaxies used for the KDEs). We use these $q$ values to estimate the lensing-bias contribution to the shear power spectra (Figure \ref{fig:red_mag_Cl}, dashed lines), and the shear correlation functions (Figure \ref{fig:red_mag_xi}, dashed lines), for the redshift binning used in the DES-SV analysis. The lensing-bias correction has the same scale dependence and similar magnitude to the reduced-shear correction, but the negative values of $q_i$ make it negative, partially cancelling out the reduced-shear correction.

We now turn to the DES Y5 forecast to demonstrate the importance of accounting for the reduced shear and lensing bias. We perform simulated likelihood analyses where either the reduced shear correction or lensing-bias correction is used to generate the `simulated' data vector from theory, but not included in the modelling during parameter estimation. For the lensing bias, we assume $q_i=-1$ for all redshift bins for simplicity (but the procedure outline in \sect{sec:lensing_bias} could be used with the DES Y5 data in order to estimate the $q_i$). \fig{fig:y5} shows the shift in the contours in the $(A,\eta_0)$ plane, due to ignoring either the reduced shear or lensing-bias. In the left panel, the case where no lensing systematics (i.e. the nuisance parameters used in the analysis of \sect{sec:SV} accounting for multiplicative shear bias, photometric redshift bias, or intrinsic alignments) are marginalized over, the shifts in the halo model parameters from ignoring these effects are significant. For the reduced shear case (green filled contour), we find $\ctruth = 97.4\%$, implying that the true values of $(A,\eta_0)$ would be excluded with 97.4\% confidence if the reduced shear correction were ignored. As expected, the shift due to ignoring lensing-bias is almost identical, but in the opposite direction. However, the shifts in the contours are still small compared to the differences between the different OWLS simulations in the $(A,\eta_0)$ plane, so we concluded that marginalising over any uncertainty (due to e.g. imperfect knowledge of the bispectrum, or the survey selection function) in the reduced shear or lensing bias corrections will not significantly reduce the power of DES Y5 to differentiate between the OWLS models used here. 

For the right panel of \fig{fig:y5}, the weak lensing nuisance parameters are now marginalised over, increasing the size of the contours, and reducing the significance of the M+15 parameter shifts. In this case, when ignoring the reduced shear, we find $\ctruth = 43.6\%$, so the recovered parameters are within `$1\sigma$' of the truth.

\subsection{Blend exclusion bias: Estimates using BCC-UFig}\label{sec:ufig}

Estimating the shear of a noisy, PSF-convolved galaxy is a notoriously difficult problem (see e.g. \citet{great3handbook}). The difficulty is further increased if the galaxy has a closely neighbouring object, since the shear estimation is likely to be disrupted by the contaminating light from the neighbour. We can categorise objects as \textit{blended} if they overlap at a particular isophotal level, for example \sex\ identifies objects by first finding groups of contiguous pixels above some detection threshold, and then deciding how many objects to split these pixels into (this decision is part of the \textit{deblending} process). If that number of objects is more than one, then these objects will be flagged as blended objects. Shape or photometry estimates (required for photo-z estimation) from these objects should be used with caution. Indeed in the DES-SV shape catalogues \citep{jarvis15}, we excluded any objects that \sex\ judged to be blended.

\citet{hartlap11} realised that this exclusion of blended objects produces a selection bias by the following mechanism: Blended objects are more likely to be in crowded regions of the sky (e.g. along the same line-of-sight as a cluster), and these crowded regions will have higher convergence than average (e.g. because of the aforementioned cluster). Therefore by excluding blended objects, we are under-sampling the higher convergence regions of the sky, compared to the less-crowded, lower convergence regions. Thus we'll underestimate the shear two-point signal, especially on small scales, where sensitivity to those crowded, high convergence regions is highest. We call this effect \textit{blend-exclusion bias}. \citet{hartlap11} estimated the magnitude of this effect by starting with a mock weak lensing catalogue (produced from ray-traced N-body simulations), and cutting out galaxies based on various criteria; for example, they apply what they call the ``FIX'' criterion, where if a pair of galaxies is separated by less than some angle $\theta_{\textrm{FIX}}$, they exclude one of those galaxies. For $\theta_{\textrm{FIX}}=2''(5'')$, they find a $-1(-2)\%$ bias in $\xip(\theta=\SI{1}{arcmin})$, and a $-2(-7)\%$ bias in $\xim(\theta=\SI{1}{arcmin})$.

These sorts of criteria give a useful indication of the expected bias; however, on real data, the criteria we use for deciding whether to use a galaxy are often not so well defined. As explained above, in the DES-SV analyses (e.g. \citealt{jarvis15}, \citealt{Be15}, \citetalias{DES15}) \sex\ was used to decide whether a galaxy is blended, and the behaviour of \sex\ is dependent on the details of the images, for example the PSF, the noise levels, and the distribution of galaxy fluxes and sizes. These details are not captured in the approach taken by \citet{hartlap11}, since they do not simulate survey images. The approach we take  uses the BCC-UFig image simulations \citep{chang15}, which allows investigation of the behaviour of the same selections we use on the real data. The BCC-UFig image simulations start with a cosmological mock galaxy simulation (the Blind Cosmology Challenge (BCC), \citealt{busha13}), with lensing information from ray-tracing \citep{becker13}. This is used as input to an image generator (the Ultra Fast Image Generator (UFig), \citealt{berge13,bruderer16}) that produces images with properties like noise levels and PSF well-matched to DES data (see \citet{chang15} and \citet{leistedt15}). The BCC-UFig catalogues are then produced by running \sex\ on these simulated images, with a configuration designed to match that run on the DES SV data by the DES data management pipeline.

\begin{figure*}
\includegraphics[width=\linewidth, trim=0cm 0cm 0cm 0cm, clip=true]{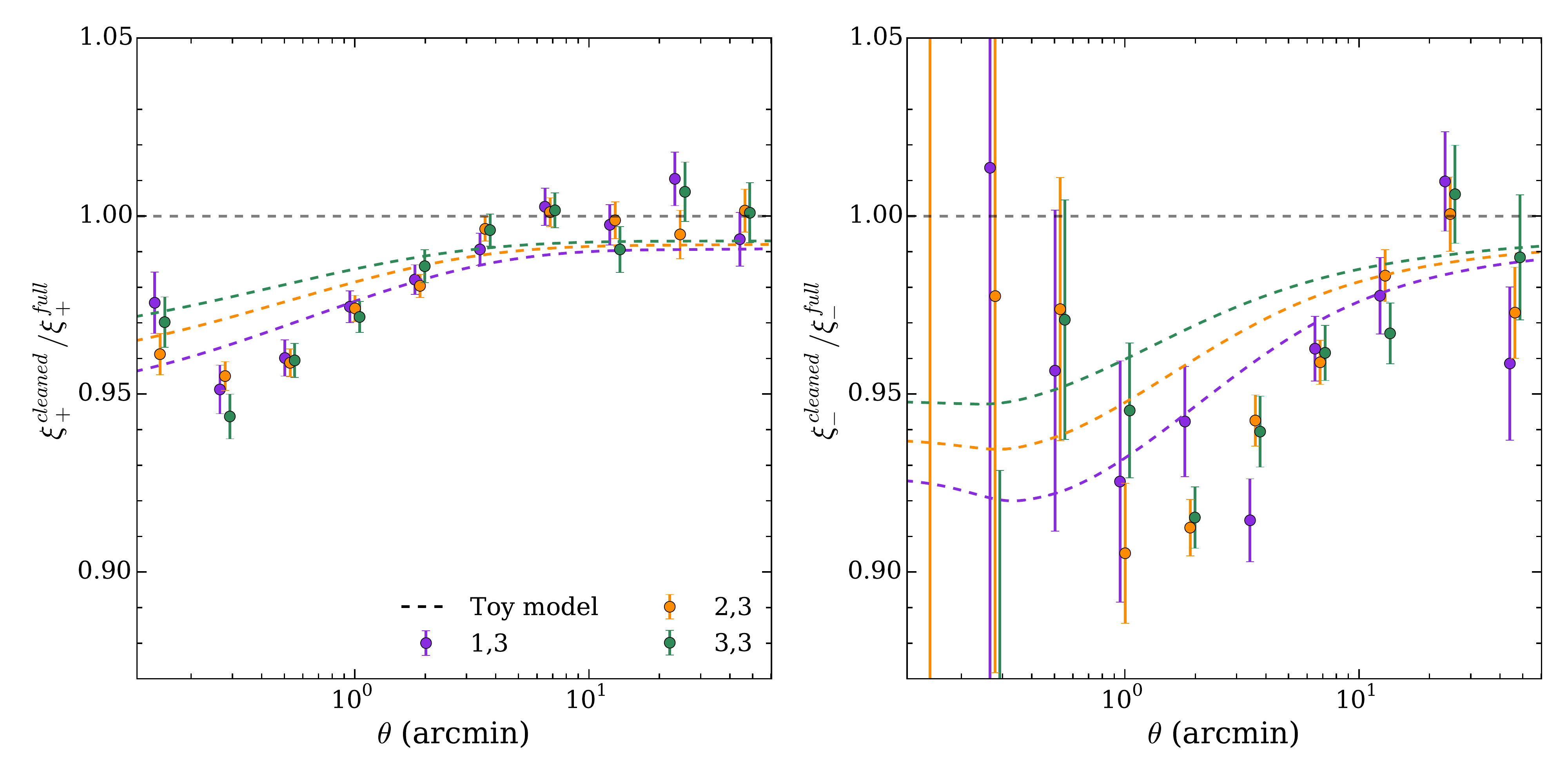}
\caption[A measurement of blend-exclusion bias using BCC-UFig.]{A measurement of blend-exclusion bias using BCC-UFig. The ratio of the shear correlation functions estimated from the BCC-UFig simulations after removing \sex\ blends, to the correlation functions using the full galaxy sample. The two samples were weighted to have the same redshift distributions, and the true input shears to the simulations were used to calculate $\xipm$, in order to isolate the selection effect. For clarity only correlations with the highest redshift bin (bin 3) are shown. The dashed lines show the prediction of the toy model described in \sect{sec:ufig}}
\label{fig:xi_ufig}
\end{figure*}

We estimate the size of the blend-exclusion bias as follows.
\begin{enumerate}
\item{We start with the DES-SV shape catalogue split into the 3 redshift bins presented in \citet{Be15} and reweight the BCC-UFig catalogue to have the same observed magnitude, size and redshift distribution. We use the $i-$band \sex~quantities MAG\_AUTO and FLUX\_RADIUS to match the magnitude and size distributions since we have these for both the SV data, and the BCC-UFig catalogues. We call this re-weighted catalogue the `full' UFig catalogue.}
\item{We measure the shear correlation functions $\xipm(\theta)$ from the full UFig catalogue, using the true input shears to the simulation. We use the true input shears, since the aim here is to isolate the selection bias, rather than study any other shape measurement biases. We call this signal $\xipm^{\textrm{full}}(\theta)$.}
\item{We then impose a cut on the \sex~flag value in the full UFig catalogue that removes blended objects or those with bright, close neighbours (around 15\% of the objects). This is the same cut that was applied to the DES-SV shape catalogues for weak lensing analyses. We re-weight the resulting catalogue to have the same redshift distribution as the full UFig catalogue, and call this the `cleaned' UFig catalogue. We measure the shear correlation functions from the cleaned UFig catalogue, and call this signal $\xipm^{\textrm{cleaned}}(\theta)$. Then the fractional bias is $\xipm^{\textrm{cleaned}}(\theta)/\xipm^{\textrm{full}}(\theta)-1$.}
\end{enumerate}
The ratio $\xipm^{\textrm{cleaned}}(\theta)/\xipm^{\textrm{full}}(\theta)$ is plotted in \fig{fig:xi_ufig}. We show only correlations with the highest redshift bin for clarity, but there is no clear redshift dependence of the bias. For $\xip$, the bias reaches $\sim3\%$ at 1 arcmin, while for $\xim$, the bias reaches this level at 10-20 arcminutes. Thus the effect is of the same order as found in \citet{hartlap11} and is similar in magnitude and scale-dependence to the reduced shear and lensing-bias effects. Thus we conclude that the blend-exclusion bias will produce a similar level of bias in the inferred M+15 halo model parameters as the reduced shear and lensing bias.

This similarity is perhaps not surprising, since this blend-exclusion bias can be thought of as a form of source-lens clustering whereby the exclusion of blended objects produces changes in the source galaxy density that are correlated with the density field; this is also the result of the lensing bias described in \sect{sec:lensing_bias}.  Motivated by this similarity, we formulate a toy model for the blend-exclusion bias.
In this toy model, we assume the probability of a galaxy at $\vectheta$ being blended depends only on the amount of light from neighbours in that area of sky. This can be quantified as the excess flux density (above the mean flux density), which we will call $\kflux(\vectheta)$. 
Consider the contribution to $\kflux(\vectheta)$ from a comoving volume element $\dx{V}$ at comoving distance $\chi$. The contribution to the excess flux in area element $\dx{\Omega}$ is
\be
\Delta \kflux(\vectheta,\chi) \dx{\chi} \dx{\Omega} = \frac{\delta_{L}(\vectheta,\chi)}{4\pi d_L(\chi)^2} \dx{V}
\ee
where $d_L(\chi)$ is the luminosity distance. $\delta_{L}(\theta,\chi)$ is the comoving volume luminosity overdensity at $(\theta,\chi)$, given by
\be
\delta_{L}(\vectheta,\chi) = \frac{L(\vectheta,\chi)-\mean{L}(\chi)}{\mean{L}(\chi)}
\ee
where $L(\vectheta,\chi)$ is the luminosity density at $(\vectheta,\chi)$ and $\mean{L}(\chi)$ is the mean luminosity density at comoving distance $\chi$. The comoving volume element can be replaced using $\dx{V} = \chi^2 \dx{\chi} \dx{\Omega}$, yielding
\be
\Delta \kflux(\vectheta,\chi) \dx{\chi} = \frac{\delta_{L}(\vectheta,\chi)}{4\pi d_L(\chi)^2} \chi^2 \dx{\chi}.
\ee
We make the assumption that the luminosity overdensity $\delta_{L}(\theta,\chi)$ is proportional to the matter overdensity $\delta(\theta,\chi)$. This would be the case if galaxies did not evolve with redshift, and had luminosity-independent bias (hence we call this a toy model!). Then
\be
\Delta \kflux(\vectheta,\chi) \dx{\chi} \propto \frac{\chi^2\delta(\vectheta,\chi)}{d_L(\chi)^2} \dx{\chi}
\ee
and
\be
\kflux(\vectheta) \propto \int \dx{\chi} \frac{\chi^2\delta(\vectheta,\chi)}{d_L(\chi)^2}. \label{eq:kflux_final}
\ee
We assume that the observed galaxy overdensity (i.e. the fractional excess in galaxy number density) due to blending is proportional to the excess flux density $\kflux(\vectheta)$, so using \eqn{eq:kflux_final},
\be
\dobs^{\text{blend}}(\vectheta) = \alpha \kflux(\vectheta) = \alpha \int \dx{\chi} \frac{\chi^2\delta(\vectheta,\chi)}{d_L(\chi)^2}
\ee
where $\alpha$ is a constant of proportionality, which we expect to be negative, since an excess in flux density should lead to more blending, and so a negative $\dobs^{\text{blend}}$.

Like the convergence, $\dobs^{\text{blend}}(\vectheta)$ is a projection in $\chi$ of the matter overdensity, $\delta$, but with a kernel 
\be
W'(\chi) = \frac{\alpha\chi^2}{d_L^2(\chi)} = \frac{\alpha\chi^2}{(1+z(\chi))^2 D_A^2(\chi)}
\ee
instead of the lensing kernel. So the effect on the shear power spectrum can be calculated in exactly the same way as the reduced shear correction, but replacing  the lensing kernels $[W_i(\chi)+W_j(\chi)]$ in equations \ref{eq:cl_red1} and \ref{eq:cl_red2}, with $2W'(\chi)$. The dashed lines in Figure \ref{fig:xi_ufig} show the prediction of this toy model, with 
$\alpha=-0.1$ showing qualitative agreement with the measurement from BCC-UFIG.

It worth noting finally that the magnitude of this selection bias (and indeed the lensing bias described in \sect{sec:lensing_bias}) will depend on the estimator used for the two-point cosmic shear signal. For example a pixel-based estimator (i.e. where the mean shear is calculated in pixels on the sky, and then these mean values are used in the two-point statistic) may be less susceptible to biases that arise from variations in the source density. However, if the pixels are weighted by the number of galaxies in each pixel, to approximate inverse-variance weighting, then in the small pixel limit, the pixel estimator estimator will be equivalent to the estimators used here.

\subsection{Intrinsic Alignments}\label{sec:IAs}

The observed intrinsic alignments of bright red galaxies (see e.g. \citealt{singh15}) on linear and mildy nonlinear scales are well described by theoretical models that assume \textit{tidal alignment}, in which the galaxy ellipticity is assumed to align with the local tidal field. The simplest of these is the linear alignment (LA) model \citep{catelan01,HS04}, in which the alignment is assumed to be linear in the linear tidal field, which leads to an alignment power spectrum that depends on the linear matter power spectrum. The LA model has only one free parameter, an amplitude $A_{\rm{IA}}$ that is of order unity (this is just called `$A$' in \citetalias{DES15}). A popular variation is the nonlinear linear alignment (NLA) model, which was introduced by \citet{BK07}, who replaced the linear matter power spectrum with the nonlinear matter power spectrum; this model has been more successful than the LA model in fitting observations on mildly nonlinear scales (e.g. \citet{joachimi10}), despite the fact that it does not include all nonlinear corrections in a consistent way. \citet{blazek15} systematically include nonlinear corrections (at one-loop order in perturbation theory) to the linear alignment model, producing a model that provides a further improved fit in the mildly nonlinear regime. Meanwhile, it is commonly assumed that the intrinsic alignments of spiral galaxies, which are primarily angular-momentum supported, are better described by theories based on tidal torquing \citet{white84}; these are also known as quadratic alignment models \citep{crittenden01,mackey02,HS04}.  Blazek et al. (in prep.) propose a perturbative model for populations of mixed galaxy-type that consistently includes both tidal alignment and tidal torque-type contributions.
Halo model-based intrinsic alignment models (see e.g. \citealt{schneider10}) are likely to be more successful in the fully nonlinear 1-halo regime. A detailed study of intrinsic alignments on nonlinear scales is beyond the scope of this work; we perform a simple test to gauge the order of the uncertainty in this section.

We use the difference between the LA model and the NLA model as a proxy for the uncertainty in the behaviour of intrinsic alignments on nonlinear scales. The green contours in \fig{fig:SV} shows the DES-SV constraints on the halo model parameters when the LA model is assumed rather than the fiducial NLA model (purple filled/outlined contours). The is a $\sim 1\sigma$ shift in the contours, towards the low $A$ favoured by the `AGN' model. This can be understood as follows: the wide redshift binning means that the dominant affect of intrinsic alignments is the negative `GI' term. The NLA model therefore produces a larger negative contribution at small scales than the LA term. When the LA model is assumed, a lower $A$ (leading to reduced halo concentration, and thus a reduced small-scale cosmic shear signal), is required to fit the observed signal.

\begin{figure}
\includegraphics[width=\linewidth, trim=0cm 0cm 0cm 0cm, clip=true]{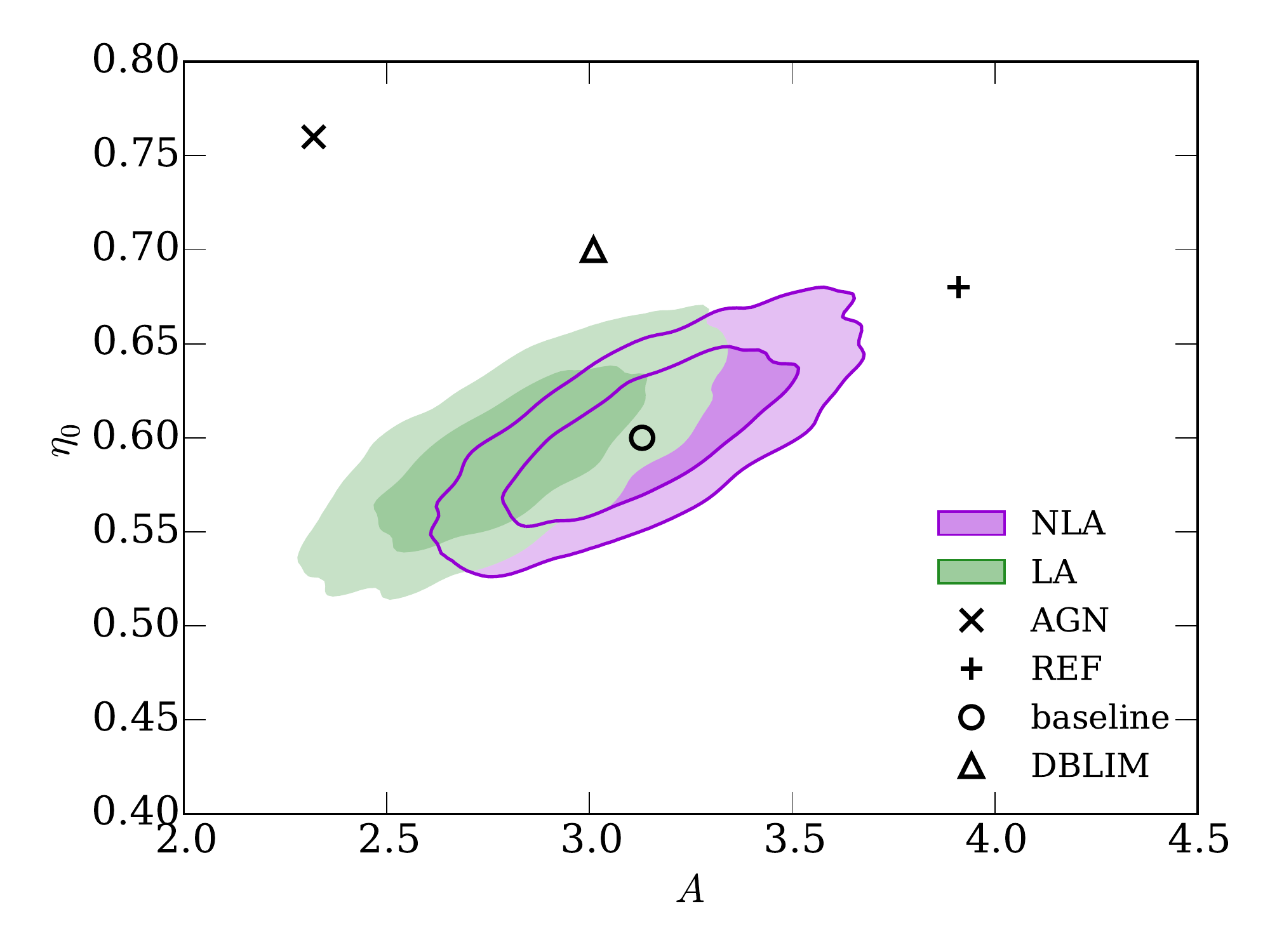}
\caption{Expected constraints from DES Y5 on M+15 halo model parameters given different assumptions about intrinsic alignments. The purple (filled and lined) contour is the same as in the right panel of \fig{fig:y5}. For the green filled contour, the linear alignment model is used to fit the simulated data vector, instead of the the NLA model which was used to generate the simulated data vector (with $A_{\rm{IA}}=0.5$). This results in biased recovery of the M+15 halo model parameters.}
\label{fig:IA}
\end{figure}

A similar shift in contours is observed for the Y5 forecast, shown as the green contours in \fig{fig:IA}. In this case the `simulated' data vector uses the NLA model, with $A_{\rm{IA}} = 0.5$, with the green contours resulting from fitting this data vector using the LA model. In this case $\ctruth=82.8\%$ so the shift is significant.
One could imagine marginalising over extra nuisance parameters to account for the uncertainty in the intrinsic alignment modelling on nonlinear scales. As a simple example, we implement an intrinsic alignment model that is a mixture of the linear alignment and NLA models, with the amount of nonlinearity determined by a free parameter $\alpha_{\rm{nl}}$, such that the intrinsic ellipticity power spectrum, $P_{\rm{II}}(k,z)$ becomes
\be
P_{\rm{II}}(k,z) = \alpha_{\rm{nl}} P_{\rm{II}}^{\rm{NLA}}(k,z) + (1 - \alpha_{\rm{nl}}) P_{\rm{II}}^{\rm{LA}},
\ee
and similarly for the intrinsic ellipticity-density cross spectrum $P_{\rm{GI}}$. Despite this extra flexibility, the degradation of the M+15 parameter constraints is negligible. While this particular model may not be very realistic, this result suggests that the future cosmic shear data will have the power to constrain additional nuisance parameters which account for the uncertainty of intrinsic alignments on nonlinear scales, without significantly degrading the constraints on the M+15 model parameters, and therefore models of baryonic physics.

\subsection{Degeneracy with cosmology}\label{sec:cosmo}

\begin{figure*}
\includegraphics[width=0.48\linewidth,  clip=true]{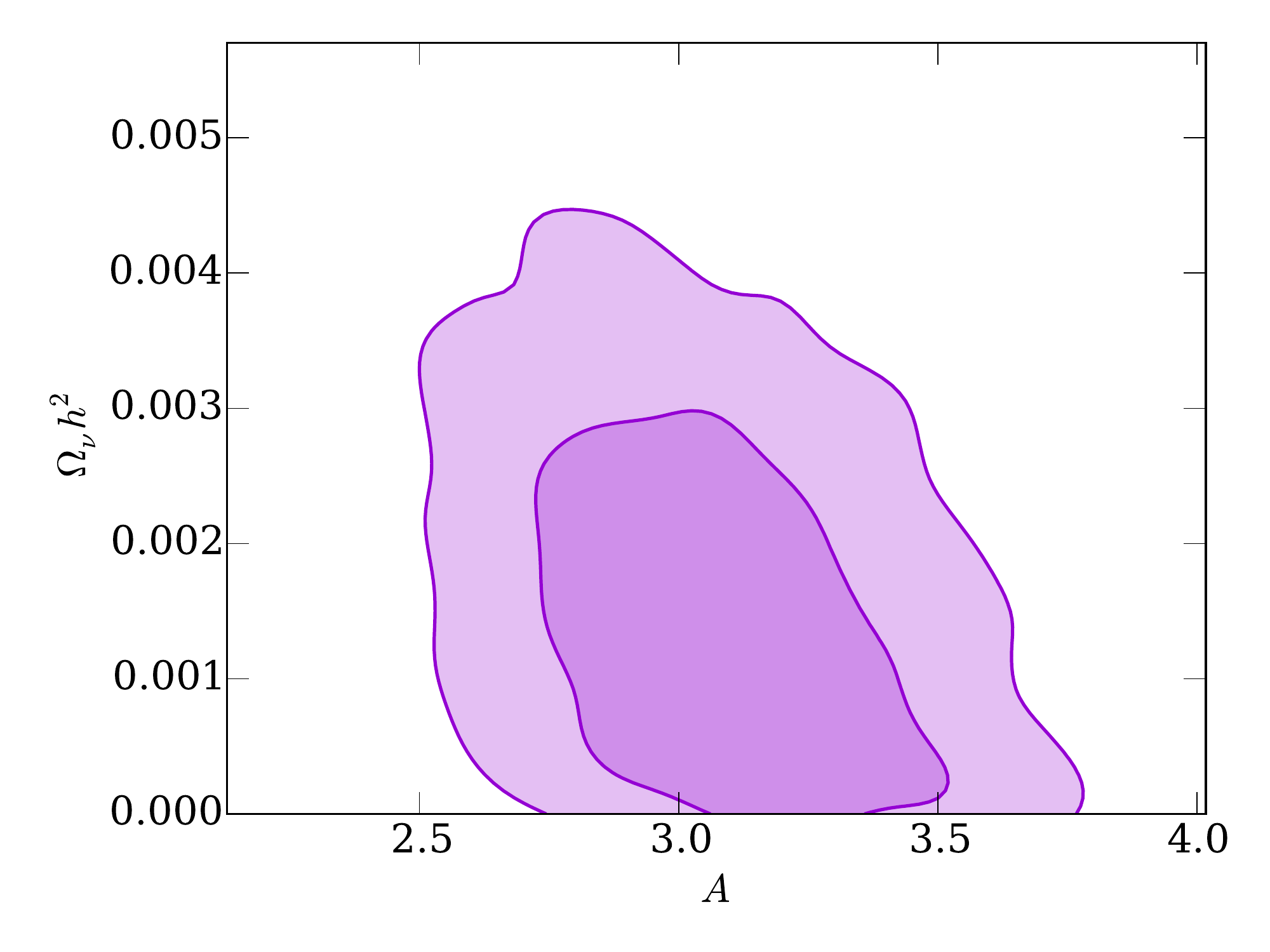}
\includegraphics[width=0.48\linewidth,  clip=true]{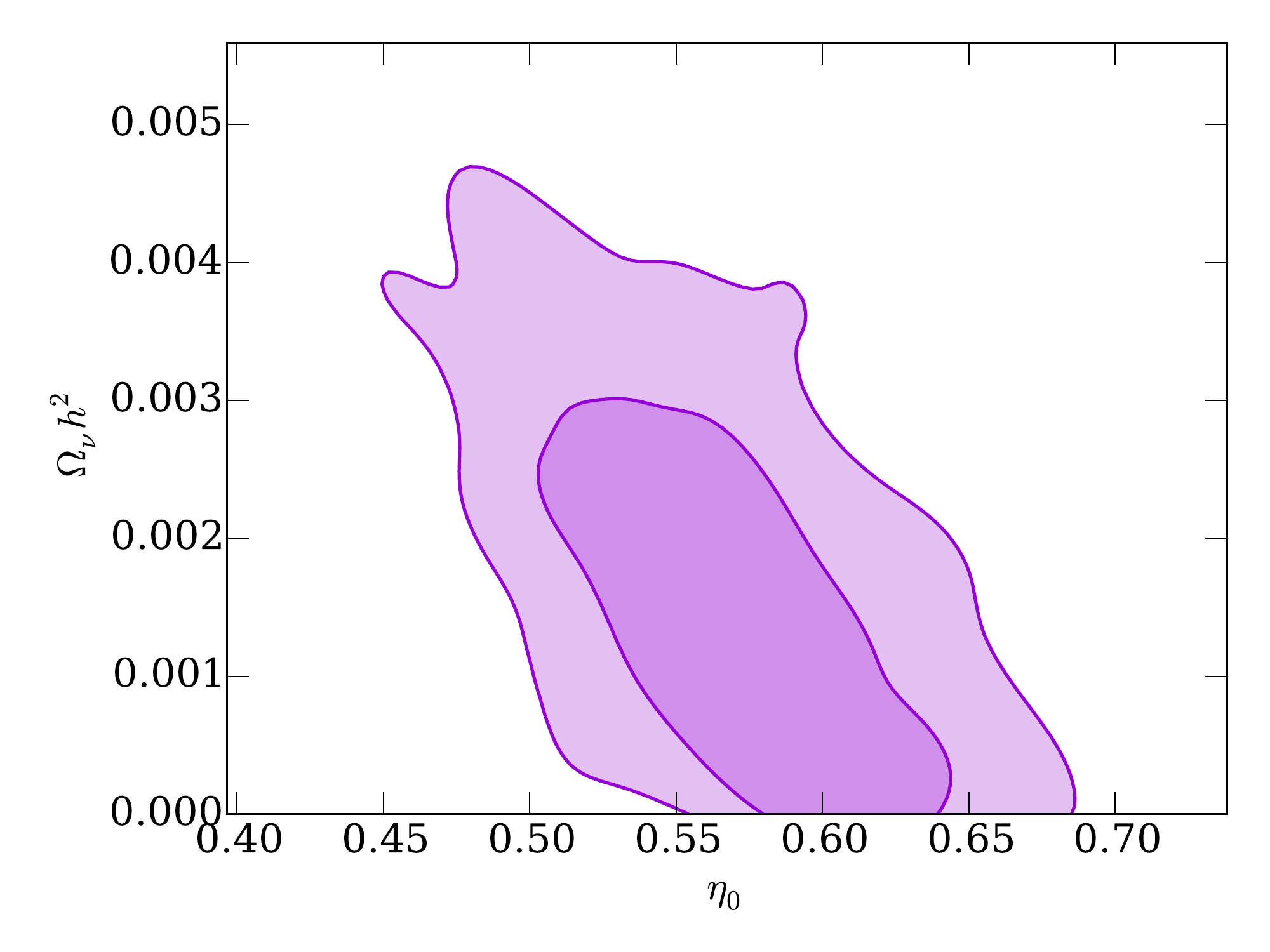}
\caption[DES Year 5 Mead+15 model forecasts.]{Forecasted degeneracy between the neutrino energy density $\Omega_\nu h^2$ and the M+15 halo model parameters, for DES Year 5 combined with \planck\ CMB constraints. For the simulated data vector, we assumed $\Omega_\nu h^2 = 6\times 10^{-4}$, approximately the minimal value allowed by solar neutrino oscillation observations \citep{fukada98}, and the halo model parameters corresponding to the baseline case ($A=3.13,\eta_0=0.60$).}
\label{fig:mnu}
\end{figure*}

Thus far, we have fixed all cosmological parameters, however, there will be some degeneracy between the halo model parameters and the cosmological parameters.
In particular, neutrino mass also produces scale dependent change in the matter power spectrum, so we expect it to have some degeneracy with baryonic feedback. \citet{natarajan14} investigate this degeneracy, concluding that one can infer biased values of the neutrino mass from cosmic shear if baryonic feedback is not accounted for. We now repeat the Y5 forecast, but allowing cosmological parameters $(\Omega_m, \Omega_b, H_0, n_s, A_s)$ to vary, while combining with \planck\ CMB constraints (specifically we use the low-$l$ TEB and high-$l$ TT likelihoods presented in \citet{planck_like15}). Note that we do not include CMB lensing information. 
There is only a small increase in the errorbars on the halo model parameters $A$ and $\eta_0$ (7\% increase in $\rtsase$). When the neutrino density $\Omega_{\nu}h^2$ is additionally marginalised over, there is a further 23\% increase in $\rtsase$. 
This degradation is due to the presence of degeneracy between the neutrino mass and the halo model parameters, as demonstrated in \fig{fig:mnu}. Hence we conclude that marginalising over cosmological parameters, including the neutrino mass, will not greatly reduce the ability of DES Y5 data to constrain baryonic effects on the matter power spectrum, when combining with \planck\ CMB data. 

\section{Discussion}\label{sec:discussion}

The small scales of cosmic shear measurements are rich in both signal-to-noise, and difficult-to-model systematic uncertainties. Baryonic effects present the largest systematic uncertainty, with $10-20\%$ deviations from the dark matter-only case on arcminute scales predicted by some hydrodynamic simulations. The prospects for gaining useful cosmological information from the small scales of cosmic shear do not look bright given these uncertainties. However, small scale cosmic shear measurements do still provide unique observational constraints on the small-scale matter clustering, since cosmic shear is the observational probe that can most directly probe the total matter distribution on small scales. These can be straightforwardly compared to e.g. the predictions from hydrodynamic simulations, or analytic models. 

We note that cosmic shear is not the only way to exploit weak lensing datasets, which (either alone or in combination with galaxy redshift surveys) can also be used for galaxy clustering measurements or probing the cross-correlation between galaxy number density and shear. \citet{viola15} have already shown the sensitivity of the latter to baryonic feedback. Furthermore, the addition of galaxy clustering and number density-shear cross-correlation information will constrain some of the systematic effects that reduce the effectiveness of cosmic shear-alone analyses, such as intrinsic alignments \citep{joachimi10b} and photometric redshift uncertainties \citep{zhang10,samuroff16}.

While current cosmic shear data has limited constraining power (such as the DES Science Verification constraints presented in \sect{sec:SV}), we have shown that information from DES 5-year data has the potential to distinguish possible baryonic scenarios, producing information that could be fed back into future hydrodynamic simulations, which in turn will hopefully improve our ability to model the small scale clustering. 
In order to make robust conclusions about baryonic physics from small scale cosmic shear however, other non-negligible systematics should be accounted for, such as the reduced shear correction, lensing bias, blend-exclusion bias, and uncertainties due to intrinsic alignment modelling. We have demonstrated that all of these effects, if not accounted for, can significantly bias the inferred small scale matter power spectrum. In particular, we have shown that these effects will bias the parameters of the \citet{mead15} halo model; however this conclusion will be true for whatever model or prescription is used to account for the uncertainties in the small-scale matter power spectrum.

While the theoretical framework for modelling the reduced shear is well established, a prediction for the matter bispectrum is required, which on nonlinear scales may also depend on the baryonic feedback.
We have demonstrated how novel image simulations can be used to estimate the effect of lensing-bias (which also requires a prediction of the matter bispectrum), and the blend-exclusion bias. Intrinsic alignment modelling on nonlinear scales is still extremely uncertain; however, we have shown the potential of future cosmic shear data to constrain uncertainty in the nonlinear intrinsic alignment modelling at the same time as the baryonic effects.
Finally, although the baryonic effects on the matter power spectrum are to some extent degenerate with the effect of massive neutrinos, we have shown that marginalising over neutrino mass does not greatly reduce the potential constraining power of DES Year 5 cosmic shear data, when it is combined with \planck\ CMB data.

\section*{Acknowledgements}

Thanks to Simon Foreman and Neil Jackson for useful discussion. Thanks to Alex Mead for help with \textsc{HMcode}, and Elisabeth Krause for producing the halo model covariance from \textsc{Cosmolike}.

We are grateful for the extraordinary contributions of our CTIO colleagues and the DECam Construction, Commissioning and Science Verification
teams in achieving the excellent instrument and telescope conditions that have made this work possible.  The success of this project also 
relies critically on the expertise and dedication of the DES Data Management group.

Funding for the DES Projects has been provided by the U.S. Department of Energy, the U.S. National Science Foundation, the Ministry of Science and Education of Spain, 
the Science and Technology Facilities Council of the United Kingdom, the Higher Education Funding Council for England, the National Center for Supercomputing 
Applications at the University of Illinois at Urbana-Champaign, the Kavli Institute of Cosmological Physics at the University of Chicago, 
the Center for Cosmology and Astro-Particle Physics at the Ohio State University,
the Mitchell Institute for Fundamental Physics and Astronomy at Texas A\&M University, Financiadora de Estudos e Projetos, 
Funda{\c c}{\~a}o Carlos Chagas Filho de Amparo {\`a} Pesquisa do Estado do Rio de Janeiro, Conselho Nacional de Desenvolvimento Cient{\'i}fico e Tecnol{\'o}gico and 
the Minist{\'e}rio da Ci{\^e}ncia, Tecnologia e Inova{\c c}{\~a}o, the Deutsche Forschungsgemeinschaft and the Collaborating Institutions in the Dark Energy Survey. 
The DES data management system is supported by the National Science Foundation under Grant Number AST-1138766.

The Collaborating Institutions are Argonne National Laboratory, the University of California at Santa Cruz, the University of Cambridge, Centro de Investigaciones En{\'e}rgeticas, 
Medioambientales y Tecnol{\'o}gicas-Madrid, the University of Chicago, University College London, the DES-Brazil Consortium, the University of Edinburgh, 
the Eidgen{\"o}ssische Technische Hochschule (ETH) Z{\"u}rich, 
Fermi National Accelerator Laboratory, the University of Illinois at Urbana-Champaign, the Institut de Ci{\`e}ncies de l'Espai (IEEC/CSIC), 
the Institut de F{\'i}sica d'Altes Energies, Lawrence Berkeley National Laboratory, the Ludwig-Maximilians Universit{\"a}t M{\"u}nchen and the associated Excellence Cluster Universe, 
the University of Michigan, the National Optical Astronomy Observatory, the University of Nottingham, The Ohio State University, the University of Pennsylvania, the University of Portsmouth, 
SLAC National Accelerator Laboratory, Stanford University, the University of Sussex, and Texas A\&M University.

The DES participants from Spanish institutions are partially supported by MINECO under grants AYA2012-39559, ESP2013-48274, FPA2013-47986, and Centro de Excelencia Severo Ochoa SEV-2012-0234 and SEV-2012-0249.
Research leading to these results has received funding from the European Research Council under the European Union Seventh Framework Programme (FP7/2007-2013) including ERC grant agreements 
 240672, 291329, and 306478.
 
 This paper has gone through internal review by the DES collaboration.




\bibliographystyle{mnras}
\bibliography{references} 

\begin{thebibliography}{}
\makeatletter
\relax
\def\mn@urlcharsother{\let\do\@makeother \do\$\do\&\do\#\do\^\do\_\do\%\do\~}
\def\mn@doi{\begingroup\mn@urlcharsother \@ifnextchar [ {\mn@doi@}
  {\mn@doi@[]}}
\def\mn@doi@[#1]#2{\def\@tempa{#1}\ifx\@tempa\@empty \href
  {http://dx.doi.org/#2} {doi:#2}\else \href {http://dx.doi.org/#2} {#1}\fi
  \endgroup}
\def\mn@eprint#1#2{\mn@eprint@#1:#2::\@nil}
\def\mn@eprint@arXiv#1{\href {http://arxiv.org/abs/#1} {{\tt arXiv:#1}}}
\def\mn@eprint@dblp#1{\href {http://dblp.uni-trier.de/rec/bibtex/#1.xml}
  {dblp:#1}}
\def\mn@eprint@#1:#2:#3:#4\@nil{\def\@tempa {#1}\def\@tempb {#2}\def\@tempc
  {#3}\ifx \@tempc \@empty \let \@tempc \@tempb \let \@tempb \@tempa \fi \ifx
  \@tempb \@empty \def\@tempb {arXiv}\fi \@ifundefined
  {mn@eprint@\@tempb}{\@tempb:\@tempc}{\expandafter \expandafter \csname
  mn@eprint@\@tempb\endcsname \expandafter{\@tempc}}}

\bibitem[\protect\citeauthoryear{{Bartelmann} \& {Schneider}}{{Bartelmann} \&
  {Schneider}}{2001}]{bartelmann01}
{Bartelmann} M.,  {Schneider} P.,  2001, \mn@doi [\physrep]
  {10.1016/S0370-1573(00)00082-X}, \href
  {http://adsabs.harvard.edu/abs/2001PhR...340..291B} {340, 291}

\bibitem[\protect\citeauthoryear{{Battye} \& {Moss}}{{Battye} \&
  {Moss}}{2014}]{battye2014}
{Battye} R.~A.,  {Moss} A.,  2014, \mn@doi [Physical Review Letters]
  {10.1103/PhysRevLett.112.051303}, \href
  {http://adsabs.harvard.edu/abs/2014PhRvL.112e1303B} {112, 051303}

\bibitem[\protect\citeauthoryear{{Becker}}{{Becker}}{2013}]{becker13}
{Becker} M.~R.,  2013, PhD thesis, The University of Chicago

\bibitem[\protect\citeauthoryear{{Becker} et~al.,}{{Becker}
  et~al.}{2015}]{Be15}
{Becker} M.~R.,  et~al., 2015, preprint, \href
  {http://adsabs.harvard.edu/abs/2015arXiv150705598B} {} (\mn@eprint {arXiv}
  {1507.05598})

\bibitem[\protect\citeauthoryear{{Berg{\'e}}, {Gamper}, {R{\'e}fr{\'e}gier}  \&
  {Amara}}{{Berg{\'e}} et~al.}{2013}]{berge13}
{Berg{\'e}} J.,  {Gamper} L.,  {R{\'e}fr{\'e}gier} A.,   {Amara} A.,  2013,
  \mn@doi [Astronomy and Computing] {10.1016/j.ascom.2013.01.001}, \href
  {http://adsabs.harvard.edu/abs/2013A%26C.....1...23B} {1, 23}

\bibitem[\protect\citeauthoryear{{Bernardeau}}{{Bernardeau}}{1998}]{bernardeau98}
{Bernardeau} F.,  1998, \aap, \href
  {http://adsabs.harvard.edu/abs/1998A%26A...338..375B} {338, 375}

\bibitem[\protect\citeauthoryear{{Bertin} \& {Arnouts}}{{Bertin} \&
  {Arnouts}}{1996}]{BertinSExtractor1996}
{Bertin} E.,  {Arnouts} S.,  1996, \aaps, \href
  {http://adsabs.harvard.edu/abs/1996A&AS..117..393B} {117, 393}

\bibitem[\protect\citeauthoryear{{Blazek}, {Vlah}  \& {Seljak}}{{Blazek}
  et~al.}{2015}]{blazek15}
{Blazek} J.,  {Vlah} Z.,   {Seljak} U.,  2015, \mn@doi [\jcap]
  {10.1088/1475-7516/2015/08/015}, \href
  {http://adsabs.harvard.edu/abs/2015JCAP...08..015B} {8, 015}

\bibitem[\protect\citeauthoryear{{Bonnett} et~al.,}{{Bonnett}
  et~al.}{2015}]{bonnett15}
{Bonnett} C.,  et~al., 2015, preprint, \href
  {http://adsabs.harvard.edu/abs/2015arXiv150705909B} {} (\mn@eprint {arXiv}
  {1507.05909})

\bibitem[\protect\citeauthoryear{{Bridle} \& {King}}{{Bridle} \&
  {King}}{2007}]{BK07}
{Bridle} S.,  {King} L.,  2007, \mn@doi [New Journal of Physics]
  {10.1088/1367-2630/9/12/444}, \href
  {http://adsabs.harvard.edu/abs/2007NJPh....9..444B} {9, 444}

\bibitem[\protect\citeauthoryear{{Bruderer}, {Chang}, {Refregier}, {Amara},
  {Berg{\'e}}  \& {Gamper}}{{Bruderer} et~al.}{2016}]{bruderer16}
{Bruderer} C.,  {Chang} C.,  {Refregier} A.,  {Amara} A.,  {Berg{\'e}} J.,
  {Gamper} L.,  2016, \mn@doi [\apj] {10.3847/0004-637X/817/1/25}, \href
  {http://adsabs.harvard.edu/abs/2016ApJ...817...25B} {817, 25}

\bibitem[\protect\citeauthoryear{{Busha}, {Wechsler}, {Becker}, {Erickson}  \&
  {Evrard}}{{Busha} et~al.}{2013}]{busha13}
{Busha} M.~T.,  {Wechsler} R.~H.,  {Becker} M.~R.,  {Erickson} B.,   {Evrard}
  A.~E.,  2013, in American Astronomical Society Meeting Abstracts \#221. p.
  341.07

\bibitem[\protect\citeauthoryear{{Casarini}, {Bonometto}, {Borgani}, {Dolag},
  {Murante}, {Mezzetti}, {Tornatore}  \& {La Vacca}}{{Casarini}
  et~al.}{2012}]{casarini12}
{Casarini} L.,  {Bonometto} S.~A.,  {Borgani} S.,  {Dolag} K.,  {Murante} G.,
  {Mezzetti} M.,  {Tornatore} L.,   {La Vacca} G.,  2012, \mn@doi [\aap]
  {10.1051/0004-6361/201118617}, \href
  {http://adsabs.harvard.edu/abs/2012A%26A...542A.126C} {542, A126}

\bibitem[\protect\citeauthoryear{{Catelan}, {Kamionkowski}  \&
  {Blandford}}{{Catelan} et~al.}{2001}]{catelan01}
{Catelan} P.,  {Kamionkowski} M.,   {Blandford} R.~D.,  2001, \mn@doi [\mnras]
  {10.1046/j.1365-8711.2001.04105.x}, \href
  {http://adsabs.harvard.edu/abs/2001MNRAS.320L...7C} {320, L7}

\bibitem[\protect\citeauthoryear{{Chang} et~al.,}{{Chang}
  et~al.}{2015}]{chang15}
{Chang} C.,  et~al., 2015, \mn@doi [\apj] {10.1088/0004-637X/801/2/73}, \href
  {http://adsabs.harvard.edu/abs/2015ApJ...801...73C} {801, 73}

\bibitem[\protect\citeauthoryear{{Clerkin} et~al.,}{{Clerkin}
  et~al.}{2016}]{clerkin16}
{Clerkin} L.,  et~al., 2016, preprint, \href
  {http://adsabs.harvard.edu/abs/2016arXiv160502036C} {} (\mn@eprint {arXiv}
  {1605.02036})

\bibitem[\protect\citeauthoryear{{Crain} et~al.,}{{Crain}
  et~al.}{2015}]{crain15}
{Crain} R.~A.,  et~al., 2015, \mn@doi [\mnras] {10.1093/mnras/stv725}, \href
  {http://adsabs.harvard.edu/abs/2015MNRAS.450.1937C} {450, 1937}

\bibitem[\protect\citeauthoryear{{Crittenden}, {Natarajan}, {Pen}  \&
  {Theuns}}{{Crittenden} et~al.}{2001}]{crittenden01}
{Crittenden} R.~G.,  {Natarajan} P.,  {Pen} U.-L.,   {Theuns} T.,  2001,
  \mn@doi [\apj] {10.1086/322370}, \href
  {http://adsabs.harvard.edu/abs/2001ApJ...559..552C} {559, 552}

\bibitem[\protect\citeauthoryear{{Cui} et~al.,}{{Cui} et~al.}{2016}]{cui16}
{Cui} W.,  et~al., 2016, \mn@doi [\mnras] {10.1093/mnras/stw603}, \href
  {http://adsabs.harvard.edu/abs/2016MNRAS.458.4052C} {458, 4052}

\bibitem[\protect\citeauthoryear{{Dodelson} \& {Schneider}}{{Dodelson} \&
  {Schneider}}{2013}]{DodelsonSchneider13}
{Dodelson} S.,  {Schneider} M.~D.,  2013, \mn@doi [\prd]
  {10.1103/PhysRevD.88.063537}, \href
  {http://adsabs.harvard.edu/abs/2013PhRvD..88f3537D} {88, 063537}

\bibitem[\protect\citeauthoryear{{Eifler}, {Krause}, {Schneider}  \&
  {Honscheid}}{{Eifler} et~al.}{2014}]{cosmolike}
{Eifler} T.,  {Krause} E.,  {Schneider} P.,   {Honscheid} K.,  2014, \mn@doi
  [\mnras] {10.1093/mnras/stu251}, \href
  {http://adsabs.harvard.edu/abs/2014MNRAS.440.1379E} {440, 1379}

\bibitem[\protect\citeauthoryear{{Eifler}, {Krause}, {Dodelson}, {Zentner},
  {Hearin}  \& {Gnedin}}{{Eifler} et~al.}{2015}]{eifler14}
{Eifler} T.,  {Krause} E.,  {Dodelson} S.,  {Zentner} A.~R.,  {Hearin} A.~P.,
  {Gnedin} N.~Y.,  2015, \mn@doi [\mnras] {10.1093/mnras/stv2000}, \href
  {http://adsabs.harvard.edu/abs/2015MNRAS.454.2451E} {454, 2451}

\bibitem[\protect\citeauthoryear{{Flaugher} et~al.,}{{Flaugher}
  et~al.}{2015}]{decam}
{Flaugher} B.,  et~al., 2015, preprint, \href
  {http://adsabs.harvard.edu/abs/2015arXiv150402900F} {} (\mn@eprint {arXiv}
  {1504.02900})

\bibitem[\protect\citeauthoryear{{Foreman}, {Becker}  \& {Wechsler}}{{Foreman}
  et~al.}{2016}]{foreman16}
{Foreman} S.,  {Becker} M.~R.,   {Wechsler} R.~H.,  2016, preprint, \href
  {http://adsabs.harvard.edu/abs/2016arXiv160509056F} {} (\mn@eprint {arXiv}
  {1605.09056})

\bibitem[\protect\citeauthoryear{{Fukuda} et~al.,}{{Fukuda}
  et~al.}{1998}]{fukada98}
{Fukuda} Y.,  et~al., 1998, \mn@doi [Physical Review Letters]
  {10.1103/PhysRevLett.81.1158}, \href
  {http://adsabs.harvard.edu/abs/1998PhRvL..81.1158F} {81, 1158}

\bibitem[\protect\citeauthoryear{{Guillet}, {Teyssier}  \& {Colombi}}{{Guillet}
  et~al.}{2010}]{guillet10}
{Guillet} T.,  {Teyssier} R.,   {Colombi} S.,  2010, \mn@doi [\mnras]
  {10.1111/j.1365-2966.2010.16466.x}, \href
  {http://adsabs.harvard.edu/abs/2010MNRAS.405..525G} {405, 525}

\bibitem[\protect\citeauthoryear{{Hamana}, {Colombi}, {Thion}, {Devriendt},
  {Mellier}  \& {Bernardeau}}{{Hamana} et~al.}{2002}]{hamana02}
{Hamana} T.,  {Colombi} S.~T.,  {Thion} A.,  {Devriendt} J.~E.~G.~T.,
  {Mellier} Y.,   {Bernardeau} F.,  2002, \mn@doi [\mnras]
  {10.1046/j.1365-8711.2002.05103.x}, \href
  {http://adsabs.harvard.edu/abs/2002MNRAS.330..365H} {330, 365}

\bibitem[\protect\citeauthoryear{{Harnois-D{\'e}raps}, {van Waerbeke}, {Viola}
  \& {Heymans}}{{Harnois-D{\'e}raps} et~al.}{2015}]{harnois14}
{Harnois-D{\'e}raps} J.,  {van Waerbeke} L.,  {Viola} M.,   {Heymans} C.,
  2015, \mn@doi [\mnras] {10.1093/mnras/stv646}, \href
  {http://adsabs.harvard.edu/abs/2015MNRAS.450.1212H} {450, 1212}

\bibitem[\protect\citeauthoryear{{Hartlap}, {Hilbert}, {Schneider}  \&
  {Hildebrandt}}{{Hartlap} et~al.}{2011}]{hartlap11}
{Hartlap} J.,  {Hilbert} S.,  {Schneider} P.,   {Hildebrandt} H.,  2011,
  \mn@doi [Astronomy \& Astrophysics] {10.1051/0004-6361/201015850}, \href
  {http://adsabs.harvard.edu/abs/2011A%26A...528A..51H} {528, A51}

\bibitem[\protect\citeauthoryear{{Hearin} \& {Zentner}}{{Hearin} \&
  {Zentner}}{2009}]{hearin09}
{Hearin} A.~P.,  {Zentner} A.~R.,  2009, \mn@doi [\jcap]
  {10.1088/1475-7516/2009/04/032}, \href
  {http://adsabs.harvard.edu/abs/2009JCAP...04..032H} {4, 032}

\bibitem[\protect\citeauthoryear{{Heitmann}, {White}, {Wagner}, {Habib}  \&
  {Higdon}}{{Heitmann} et~al.}{2010}]{heitmann10}
{Heitmann} K.,  {White} M.,  {Wagner} C.,  {Habib} S.,   {Higdon} D.,  2010,
  \mn@doi [\apj] {10.1088/0004-637X/715/1/104}, \href
  {http://adsabs.harvard.edu/abs/2010ApJ...715..104H} {715, 104}

\bibitem[\protect\citeauthoryear{{Heitmann}, {Lawrence}, {Kwan}, {Habib}  \&
  {Higdon}}{{Heitmann} et~al.}{2014}]{heitmann14}
{Heitmann} K.,  {Lawrence} E.,  {Kwan} J.,  {Habib} S.,   {Higdon} D.,  2014,
  \mn@doi [\apj] {10.1088/0004-637X/780/1/111}, \href
  {http://adsabs.harvard.edu/abs/2014ApJ...780..111H} {780, 111}

\bibitem[\protect\citeauthoryear{{Heymans} et~al.,}{{Heymans}
  et~al.}{2012}]{heymans12}
{Heymans} C.,  et~al., 2012, \mn@doi [\mnras]
  {10.1111/j.1365-2966.2012.21952.x}, \href
  {http://adsabs.harvard.edu/abs/2012MNRAS.427..146H} {427, 146}

\bibitem[\protect\citeauthoryear{{Hilbert}, {Hartlap}  \&
  {Schneider}}{{Hilbert} et~al.}{2011}]{hilbert11}
{Hilbert} S.,  {Hartlap} J.,   {Schneider} P.,  2011, \mn@doi [\aap]
  {10.1051/0004-6361/201117294}, \href
  {http://adsabs.harvard.edu/abs/2011A%26A...536A..85H} {536, A85}

\bibitem[\protect\citeauthoryear{{Hildebrandt} et~al.,}{{Hildebrandt}
  et~al.}{2016}]{hildebrandt16}
{Hildebrandt} H.,  et~al., 2016, preprint, \href
  {http://adsabs.harvard.edu/abs/2016arXiv160605338H} {} (\mn@eprint {arXiv}
  {1606.05338})

\bibitem[\protect\citeauthoryear{{Hinshaw} et~al.,}{{Hinshaw}
  et~al.}{2013}]{wmap9}
{Hinshaw} G.,  et~al., 2013, \mn@doi [\apjs] {10.1088/0067-0049/208/2/19},
  \href {http://adsabs.harvard.edu/abs/2013ApJS..208...19H} {208, 19}

\bibitem[\protect\citeauthoryear{{Hirata} \& {Seljak}}{{Hirata} \&
  {Seljak}}{2004}]{HS04}
{Hirata} C.~M.,  {Seljak} U.,  2004, \mn@doi [\prd]
  {10.1103/PhysRevD.70.063526}, \href
  {http://adsabs.harvard.edu/abs/2004PhRvD..70f3526H} {70, 063526}

\bibitem[\protect\citeauthoryear{{Huterer} \& {Takada}}{{Huterer} \&
  {Takada}}{2005}]{huterer05}
{Huterer} D.,  {Takada} M.,  2005, \mn@doi [Astroparticle Physics]
  {10.1016/j.astropartphys.2005.02.006}, \href
  {http://adsabs.harvard.edu/abs/2005APh....23..369H} {23, 369}

\bibitem[\protect\citeauthoryear{{Jarvis} et~al.,}{{Jarvis}
  et~al.}{2016}]{jarvis15}
{Jarvis} M.,  et~al., 2016, \mn@doi [\mnras] {10.1093/mnras/stw990}, \href
  {http://adsabs.harvard.edu/abs/2016MNRAS.460.2245J} {460, 2245}

\bibitem[\protect\citeauthoryear{{Jing}, {Zhang}, {Lin}, {Gao}  \&
  {Springel}}{{Jing} et~al.}{2006}]{jing06}
{Jing} Y.~P.,  {Zhang} P.,  {Lin} W.~P.,  {Gao} L.,   {Springel} V.,  2006,
  \mn@doi [\apjl] {10.1086/503547}, \href
  {http://adsabs.harvard.edu/abs/2006ApJ...640L.119J} {640, L119}

\bibitem[\protect\citeauthoryear{{Joachimi} \& {Bridle}}{{Joachimi} \&
  {Bridle}}{2010}]{joachimi10b}
{Joachimi} B.,  {Bridle} S.~L.,  2010, \mn@doi [\aap]
  {10.1051/0004-6361/200913657}, \href
  {http://adsabs.harvard.edu/abs/2010A%26A...523A...1J} {523, A1}

\bibitem[\protect\citeauthoryear{{Joachimi} \& {Schneider}}{{Joachimi} \&
  {Schneider}}{2010}]{joachimi10}
{Joachimi} B.,  {Schneider} P.,  2010, preprint, \href
  {http://adsabs.harvard.edu/abs/2010arXiv1009.2024J} {} (\mn@eprint {arXiv}
  {1009.2024})

\bibitem[\protect\citeauthoryear{{Joudaki} et~al.,}{{Joudaki}
  et~al.}{2016}]{joudaki16}
{Joudaki} S.,  et~al., 2016, preprint, \href
  {http://adsabs.harvard.edu/abs/2016arXiv160105786J} {} (\mn@eprint {arXiv}
  {1601.05786})

\bibitem[\protect\citeauthoryear{{Kitching} et~al.,}{{Kitching}
  et~al.}{2014}]{kitching14}
{Kitching} T.~D.,  et~al., 2014, preprint, \href
  {http://adsabs.harvard.edu/abs/2014arXiv1401.6842K} {} (\mn@eprint {arXiv}
  {1401.6842})

\bibitem[\protect\citeauthoryear{{Kitching}, {Verde}, {Heavens}  \&
  {Jimenez}}{{Kitching} et~al.}{2016}]{kitching16}
{Kitching} T.~D.,  {Verde} L.,  {Heavens} A.~F.,   {Jimenez} R.,  2016, \mn@doi
  [\mnras] {10.1093/mnras/stw707}, \href
  {http://adsabs.harvard.edu/abs/2016MNRAS.459..971K} {459, 971}

\bibitem[\protect\citeauthoryear{{Krause} \& {Eifler}}{{Krause} \&
  {Eifler}}{2016}]{krause16}
{Krause} E.,  {Eifler} T.,  2016, preprint, \href
  {http://adsabs.harvard.edu/abs/2016arXiv160105779K} {} (\mn@eprint {arXiv}
  {1601.05779})

\bibitem[\protect\citeauthoryear{{Krause} \& {Hirata}}{{Krause} \&
  {Hirata}}{2010}]{krause10}
{Krause} E.,  {Hirata} C.~M.,  2010, \mn@doi [\aap]
  {10.1051/0004-6361/200913524}, \href
  {http://adsabs.harvard.edu/abs/2010A%26A...523A..28K} {523, A28}

\bibitem[\protect\citeauthoryear{{Leistedt} et~al.,}{{Leistedt}
  et~al.}{2015}]{leistedt15}
{Leistedt} B.,  et~al., 2015, preprint, \href
  {http://adsabs.harvard.edu/abs/2015arXiv150705647L} {} (\mn@eprint {arXiv}
  {1507.05647})

\bibitem[\protect\citeauthoryear{{MacCrann}, {Zuntz}, {Bridle}, {Jain}  \&
  {Becker}}{{MacCrann} et~al.}{2015}]{maccrann15}
{MacCrann} N.,  {Zuntz} J.,  {Bridle} S.,  {Jain} B.,   {Becker} M.~R.,  2015,
  \mn@doi [\mnras] {10.1093/mnras/stv1154}, \href
  {http://adsabs.harvard.edu/abs/2015MNRAS.451.2877M} {451, 2877}

\bibitem[\protect\citeauthoryear{{Mackey}, {White}  \& {Kamionkowski}}{{Mackey}
  et~al.}{2002}]{mackey02}
{Mackey} J.,  {White} M.,   {Kamionkowski} M.,  2002, \mn@doi [\mnras]
  {10.1046/j.1365-8711.2002.05337.x}, \href
  {http://adsabs.harvard.edu/abs/2002MNRAS.332..788M} {332, 788}

\bibitem[\protect\citeauthoryear{{Mandelbaum} et~al.,}{{Mandelbaum}
  et~al.}{2014}]{great3handbook}
{Mandelbaum} R.,  et~al., 2014, \mn@doi [\apjs] {10.1088/0067-0049/212/1/5},
  \href {http://adsabs.harvard.edu/abs/2014ApJS..212....5M} {212, 5}

\bibitem[\protect\citeauthoryear{{Mead}, {Peacock}, {Heymans}, {Joudaki}  \&
  {Heavens}}{{Mead} et~al.}{2015}]{mead15}
{Mead} A.,  {Peacock} J.,  {Heymans} C.,  {Joudaki} S.,   {Heavens} A.,  2015,
  preprint, \href {http://adsabs.harvard.edu/abs/2015arXiv150507833M} {}
  (\mn@eprint {arXiv} {1505.07833})

\bibitem[\protect\citeauthoryear{{Natarajan}, {Zentner}, {Battaglia}  \&
  {Trac}}{{Natarajan} et~al.}{2014}]{natarajan14}
{Natarajan} A.,  {Zentner} A.~R.,  {Battaglia} N.,   {Trac} H.,  2014, \mn@doi
  [\prd] {10.1103/PhysRevD.90.063516}, \href
  {http://adsabs.harvard.edu/abs/2014PhRvD..90f3516N} {90, 063516}

\bibitem[\protect\citeauthoryear{{Navarro}, {Frenk}  \& {White}}{{Navarro}
  et~al.}{1996}]{navarro96}
{Navarro} J.~F.,  {Frenk} C.~S.,   {White} S.~D.~M.,  1996, \mn@doi [\apj]
  {10.1086/177173}, \href {http://adsabs.harvard.edu/abs/1996ApJ...462..563N}
  {462, 563}

\bibitem[\protect\citeauthoryear{{Peacock} \& {Smith}}{{Peacock} \&
  {Smith}}{2000}]{peacock00}
{Peacock} J.~A.,  {Smith} R.~E.,  2000, \mn@doi [\mnras]
  {10.1046/j.1365-8711.2000.03779.x}, \href
  {http://adsabs.harvard.edu/abs/2000MNRAS.318.1144P} {318, 1144}

\bibitem[\protect\citeauthoryear{{Planck Collaboration} et~al.,}{{Planck
  Collaboration} et~al.}{2013}]{planckcosmo13}
{Planck Collaboration} et~al., 2013, preprint, \href
  {http://adsabs.harvard.edu/abs/2013arXiv1303.5076P} {} (\mn@eprint {arXiv}
  {1303.5076})

\bibitem[\protect\citeauthoryear{{Planck Collaboration} et~al.,}{{Planck
  Collaboration} et~al.}{2015b}]{planck_like15}
{Planck Collaboration} et~al., 2015b, preprint, \href
  {http://adsabs.harvard.edu/abs/2015arXiv150702704P} {} (\mn@eprint {arXiv}
  {1507.02704})

\bibitem[\protect\citeauthoryear{{Planck Collaboration} et~al.,}{{Planck
  Collaboration} et~al.}{2015a}]{planckcosmo15}
{Planck Collaboration} et~al., 2015a, preprint, \href
  {http://adsabs.harvard.edu/abs/2015arXiv150201589P} {} (\mn@eprint {arXiv}
  {1502.01589})

\bibitem[\protect\citeauthoryear{{Rudd}, {Zentner}  \& {Kravtsov}}{{Rudd}
  et~al.}{2008}]{rudd08}
{Rudd} D.~H.,  {Zentner} A.~R.,   {Kravtsov} A.~V.,  2008, \mn@doi [\apj]
  {10.1086/523836}, \href {http://adsabs.harvard.edu/abs/2008ApJ...672...19R}
  {672, 19}

\bibitem[\protect\citeauthoryear{{Samuroff}, {Troxel}, {Bridle}, {Zuntz},
  {MacCrann}, {Krause}, {Eifler}  \& {Kirk}}{{Samuroff}
  et~al.}{2016}]{samuroff16}
{Samuroff} S.,  {Troxel} M.,  {Bridle} S.,  {Zuntz} J.,  {MacCrann} N.,
  {Krause} E.,  {Eifler} T.,   {Kirk} D.,  2016, preprint, \href
  {http://adsabs.harvard.edu/abs/2016arXiv160707910S} {} (\mn@eprint {arXiv}
  {1607.07910})

\bibitem[\protect\citeauthoryear{{Sato}, {Hamana}, {Takahashi}, {Takada},
  {Yoshida}, {Matsubara}  \& {Sugiyama}}{{Sato} et~al.}{2009}]{sato09}
{Sato} M.,  {Hamana} T.,  {Takahashi} R.,  {Takada} M.,  {Yoshida} N.,
  {Matsubara} T.,   {Sugiyama} N.,  2009, \mn@doi [\apj]
  {10.1088/0004-637X/701/2/945}, \href
  {http://adsabs.harvard.edu/abs/2009ApJ...701..945S} {701, 945}

\bibitem[\protect\citeauthoryear{{Schaye} et~al.,}{{Schaye}
  et~al.}{2010}]{schaye10}
{Schaye} J.,  et~al., 2010, \mn@doi [\mnras]
  {10.1111/j.1365-2966.2009.16029.x}, \href
  {http://adsabs.harvard.edu/abs/2010MNRAS.402.1536S} {402, 1536}

\bibitem[\protect\citeauthoryear{{Schaye} et~al.,}{{Schaye}
  et~al.}{2015}]{schaye15}
{Schaye} J.,  et~al., 2015, \mn@doi [\mnras] {10.1093/mnras/stu2058}, \href
  {http://adsabs.harvard.edu/abs/2015MNRAS.446..521S} {446, 521}

\bibitem[\protect\citeauthoryear{{Schmidt}, {Rozo}, {Dodelson}, {Hui}  \&
  {Sheldon}}{{Schmidt} et~al.}{2009}]{schmidt09}
{Schmidt} F.,  {Rozo} E.,  {Dodelson} S.,  {Hui} L.,   {Sheldon} E.,  2009,
  \mn@doi [\apj] {10.1088/0004-637X/702/1/593}, \href
  {http://adsabs.harvard.edu/abs/2009ApJ...702..593S} {702, 593}

\bibitem[\protect\citeauthoryear{{Schneider} \& {Bridle}}{{Schneider} \&
  {Bridle}}{2010}]{schneider10}
{Schneider} M.~D.,  {Bridle} S.,  2010, \mn@doi [\mnras]
  {10.1111/j.1365-2966.2009.15956.x}, \href
  {http://adsabs.harvard.edu/abs/2010MNRAS.402.2127S} {402, 2127}

\bibitem[\protect\citeauthoryear{{Scoccimarro} \& {Couchman}}{{Scoccimarro} \&
  {Couchman}}{2001}]{scoccimarro01}
{Scoccimarro} R.,  {Couchman} H.~M.~P.,  2001, \mn@doi [\mnras]
  {10.1046/j.1365-8711.2001.04281.x}, \href
  {http://adsabs.harvard.edu/abs/2001MNRAS.325.1312S} {325, 1312}

\bibitem[\protect\citeauthoryear{{Seljak}}{{Seljak}}{2000}]{seljak00}
{Seljak} U.,  2000, \mn@doi [\mnras] {10.1046/j.1365-8711.2000.03715.x}, \href
  {http://adsabs.harvard.edu/abs/2000MNRAS.318..203S} {318, 203}

\bibitem[\protect\citeauthoryear{{Semboloni}, {Hoekstra}, {Schaye}, {van
  Daalen}  \& {McCarthy}}{{Semboloni} et~al.}{2011}]{sembolini11}
{Semboloni} E.,  {Hoekstra} H.,  {Schaye} J.,  {van Daalen} M.~P.,   {McCarthy}
  I.~G.,  2011, \mn@doi [\mnras] {10.1111/j.1365-2966.2011.19385.x}, \href
  {http://adsabs.harvard.edu/abs/2011MNRAS.417.2020S} {417, 2020}

\bibitem[\protect\citeauthoryear{{Shapiro}}{{Shapiro}}{2009}]{shapiro09}
{Shapiro} C.,  2009, \mn@doi [\apj] {10.1088/0004-637X/696/1/775}, \href
  {http://adsabs.harvard.edu/abs/2009ApJ...696..775S} {696, 775}

\bibitem[\protect\citeauthoryear{{Sheldon}}{{Sheldon}}{2014}]{Sheldon2014}
{Sheldon} E.~S.,  2014, \mn@doi [\mnras] {10.1093/mnrasl/slu104}, \href
  {http://adsabs.harvard.edu/abs/2014MNRAS.444L..25S} {444, L25}

\bibitem[\protect\citeauthoryear{{Singh}, {Mandelbaum}  \& {More}}{{Singh}
  et~al.}{2015}]{singh15}
{Singh} S.,  {Mandelbaum} R.,   {More} S.,  2015, \mn@doi [\mnras]
  {10.1093/mnras/stv778}, \href
  {http://adsabs.harvard.edu/abs/2015MNRAS.450.2195S} {450, 2195}

\bibitem[\protect\citeauthoryear{{Smith} et~al.,}{{Smith}
  et~al.}{2003}]{smith03}
{Smith} R.~E.,  et~al., 2003, \mn@doi [\mnras]
  {10.1046/j.1365-8711.2003.06503.x}, \href
  {http://adsabs.harvard.edu/abs/2003MNRAS.341.1311S} {341, 1311}

\bibitem[\protect\citeauthoryear{{Suchyta} et~al.,}{{Suchyta}
  et~al.}{2016}]{suchyta16}
{Suchyta} E.,  et~al., 2016, \mn@doi [\mnras] {10.1093/mnras/stv2953}, \href
  {http://adsabs.harvard.edu/abs/2016MNRAS.457..786S} {457, 786}

\bibitem[\protect\citeauthoryear{{Takada} \& {Hu}}{{Takada} \&
  {Hu}}{2013}]{takada13}
{Takada} M.,  {Hu} W.,  2013, \mn@doi [\prd] {10.1103/PhysRevD.87.123504},
  \href {http://adsabs.harvard.edu/abs/2013PhRvD..87l3504T} {87, 123504}

\bibitem[\protect\citeauthoryear{{Takahashi}, {Oguri}, {Sato}  \&
  {Hamana}}{{Takahashi} et~al.}{2011}]{takahashi11}
{Takahashi} R.,  {Oguri} M.,  {Sato} M.,   {Hamana} T.,  2011, \mn@doi [\apj]
  {10.1088/0004-637X/742/1/15}, \href
  {http://adsabs.harvard.edu/abs/2011ApJ...742...15T} {742, 15}

\bibitem[\protect\citeauthoryear{{Takahashi}, {Sato}, {Nishimichi}, {Taruya}
  \& {Oguri}}{{Takahashi} et~al.}{2012}]{takahashi2012}
{Takahashi} R.,  {Sato} M.,  {Nishimichi} T.,  {Taruya} A.,   {Oguri} M.,
  2012, \mn@doi [\apj] {10.1088/0004-637X/761/2/152}, \href
  {http://adsabs.harvard.edu/abs/2012ApJ...761..152T} {761, 152}

\bibitem[\protect\citeauthoryear{{Taruya}, {Takada}, {Hamana}, {Kayo}  \&
  {Futamase}}{{Taruya} et~al.}{2002}]{taruya02}
{Taruya} A.,  {Takada} M.,  {Hamana} T.,  {Kayo} I.,   {Futamase} T.,  2002,
  \mn@doi [\apj] {10.1086/340048}, \href
  {http://adsabs.harvard.edu/abs/2002ApJ...571..638T} {571, 638}

\bibitem[\protect\citeauthoryear{{Taylor}, {Joachimi}  \& {Kitching}}{{Taylor}
  et~al.}{2013}]{taylor2013}
{Taylor} A.,  {Joachimi} B.,   {Kitching} T.,  2013, \mn@doi [\mnras]
  {10.1093/mnras/stt270}, \href
  {http://adsabs.harvard.edu/abs/2013MNRAS.432.1928T} {432, 1928}

\bibitem[\protect\citeauthoryear{{The Dark Energy Survey Collaboration}
  et~al.,}{{The Dark Energy Survey Collaboration} et~al.}{2016}]{DES15}
{The Dark Energy Survey Collaboration} et~al., 2016, \mn@doi [Phys. Rev. D]
  {10.1103/PhysRevD.94.022001}, 94, 022001

\bibitem[\protect\citeauthoryear{{Valageas}}{{Valageas}}{2014}]{valegas14}
{Valageas} P.,  2014, \mn@doi [\aap] {10.1051/0004-6361/201322146}, \href
  {http://adsabs.harvard.edu/abs/2014A%26A...561A..53V} {561, A53}

\bibitem[\protect\citeauthoryear{{Viola} et~al.,}{{Viola}
  et~al.}{2015}]{viola15}
{Viola} M.,  et~al., 2015, \mn@doi [\mnras] {10.1093/mnras/stv1447}, \href
  {http://adsabs.harvard.edu/abs/2015MNRAS.452.3529V} {452, 3529}

\bibitem[\protect\citeauthoryear{{White}}{{White}}{1984}]{white84}
{White} S.~D.~M.,  1984, \mn@doi [\apj] {10.1086/162573}, \href
  {http://adsabs.harvard.edu/abs/1984ApJ...286...38W} {286, 38}

\bibitem[\protect\citeauthoryear{{White}}{{White}}{2004}]{white04}
{White} M.,  2004, \mn@doi [Astroparticle Physics]
  {10.1016/j.astropartphys.2004.06.001}, \href
  {http://adsabs.harvard.edu/abs/2004APh....22..211W} {22, 211}

\bibitem[\protect\citeauthoryear{{Zentner}, {Rudd}  \& {Hu}}{{Zentner}
  et~al.}{2008}]{zentner08}
{Zentner} A.~R.,  {Rudd} D.~H.,   {Hu} W.,  2008, \mn@doi [\prd]
  {10.1103/PhysRevD.77.043507}, \href
  {http://adsabs.harvard.edu/abs/2008PhRvD..77d3507Z} {77, 043507}

\bibitem[\protect\citeauthoryear{{Zhan} \& {Knox}}{{Zhan} \&
  {Knox}}{2004}]{zhan04}
{Zhan} H.,  {Knox} L.,  2004, \mn@doi [\apjl] {10.1086/426712}, \href
  {http://adsabs.harvard.edu/abs/2004ApJ...616L..75Z} {616, L75}

\bibitem[\protect\citeauthoryear{{Zhang}, {Pen}  \& {Bernstein}}{{Zhang}
  et~al.}{2010}]{zhang10}
{Zhang} P.,  {Pen} U.-L.,   {Bernstein} G.,  2010, \mn@doi [\mnras]
  {10.1111/j.1365-2966.2010.16445.x}, \href
  {http://adsabs.harvard.edu/abs/2010MNRAS.405..359Z} {405, 359}

\bibitem[\protect\citeauthoryear{Zuntz et~al.,}{Zuntz et~al.}{2015}]{cosmosis}
Zuntz J.,  et~al., 2015, \mn@doi [Astronomy and Computing]
  {http://dx.doi.org/10.1016/j.ascom.2015.05.005}, 12, 45

\bibitem[\protect\citeauthoryear{{van Daalen}, {Schaye}, {Booth}  \& {Dalla
  Vecchia}}{{van Daalen} et~al.}{2011}]{vandalen11}
{van Daalen} M.~P.,  {Schaye} J.,  {Booth} C.~M.,   {Dalla Vecchia} C.,  2011,
  \mn@doi [\mnras] {10.1111/j.1365-2966.2011.18981.x}, \href
  {http://adsabs.harvard.edu/abs/2011MNRAS.415.3649V} {415, 3649}

\makeatother
\end{thebibliography}



\appendix

\section{Third order corrections to shear-shear correlations}
\label{app:shear3}


In cosmic shear, we attempt to measure the two-point correlation of the shear, possibly between two different redshift bins $i$ and $j$
\be
\xi_{i,j} = \langle \gamma_i(\vec{x})\gamma_j(\vec{x'}) \rangle.
\ee
Contributions to shear two-point correlation at third order in the density field arise from two effects
\begin{enumerate}
\item{We observe the \textit{reduced shear}, 
\begin{equation}
g(\vec{x})=\frac{\gamma(\vec{x})}{1-\kappa(\vec{x})}\approx(1+\kappa(\vec{x}))\gamma(\vec{x}).
\end{equation}}
\item{We can only estimate the shear at positions of galaxies. So any statistic (e.g. the mean shear or $\xipm$) estimated from the measured shears will effectively be using the galaxy number density-weighted reduced shear:
\begin{align}
g^{obs}(\vec{x}) &= (1 + \dobs(\vec{x}))g(\vec{x})) \\
&= (1 + \dobs(\vec{x}))(1+\kappa(\vec{x}))\gamma(\vec{x})
\label{eq:gobs}
\end{align}
where $\dobs(\vec{x})$ is the observed overdensity in galaxy number at position $\vec{x}$. This observed overdensity can be due to a true change in the number density of galaxies at $\vec{x}$ (for example due to the presence of a cluster), or due to a change in the \textit{observable} number density due to lensing magnification (for example due to the presence of a cluster at lower redshift). The first effect leads to \textit{source-lens clustering} \citep{bernardeau98,hamana02}  and the second leads to \textit{lensing-bias} \citep{schmidt09}}.
\end{enumerate}

We start with the expression from \citet{schmidt09} for the expectation of the standard $\xipm$ estimator.
\begin{equation}
\langle\xi_{ij}^{\obs}\rangle = \left\langle\frac{ g_i^{\obs}(\vec{x})g_j^{\obs}(\vec{x'})}{1 + 2\overline{\dobs} + \widehat{\dobs\dobs}}\right\rangle
\end{equation}
where $\overline{\dobs}$ is the mean observed galaxy overdensity across the survey (negligible for a wide survey), $\widehat{\dobs\dobs}$ is a mean product of overdensities smoothed over the bin width ($\xi_{gg}(\theta)$ in the limit of an infinite survey and narrow bin). Substituting for $g_i^{\obs}$ from \eqn{eq:gobs}, the terms up to third order in $\gamma$, $\kappa$ or $\delta$ are
\begin{align}
\langle\xi_{ij}^{\obs}\rangle = &\langle\gamma_i(\vec{x})\gamma_j(\vec{x'})\rangle\notag\\
+ &\langle\kappa(\vec{x})\gamma_i(\vec{x})\gamma_j(\vec{x'})\rangle + \langle\gamma_i(\vec{x})\kappa(\vec{x'})\gamma_j(\vec{x'})\rangle \\
+ &\langle\dobs(\vec{x})\gamma_i(\vec{x})\gamma_j(\vec{x'})\rangle + \langle\gamma_i(\vec{x})\dobs(\vec{x'})\gamma_j(\vec{x'})\rangle\notag.
\end{align}
The first line is the `true' shear-shear signal. The second line is the reduced shear contribution, which is only zero if the convergence $\kappa$ is not correlated with the shear at a given point on the sky, which is unlikely, since they are sourced by the same structure. The third line is the source-lens clustering (including `lensing bias', since magnification contributes to $\dobs$). This would be zero if there was no correlation between the source galaxy overdensity and the shear at a given point on the sky e.g. if source galaxies were randomly distributed. 

It's more convenient to compute these term in Fourier space, as
\begin{equation}
\begin{split}
\langle\gamma^{\obs}_i(\vec{l},\chi)\gamma^{\obs,*}_j(\vec{l''},\chi')\rangle=&\
\langle\gamma_i(\vec{l},\chi)\gamma_j^*(\vec{l''},\chi')\rangle \\&+
R_{ij} + R_{ji} + S_{ij} + S_{ji}
\end{split}
\label{eqn:fourier_start}
\end{equation}
where
\begin{align}
R_{ij} &= \langle(\kappa_i\gamma_i)(\vec{l},\chi)\gamma_j^*(\vec{l''},\chi'))\rangle\\
R_{ji} &= \langle\gamma_i(\vec{l},\chi)(\kappa_j\gamma_j)^*(\vec{l''},\chi')\rangle\\
S_{ij} &= \langle(\dobsi\gamma_i)(\vec{l},\chi)\gamma_j^*(\vec{l''},\chi'))\rangle\\
S_{ji} &= \langle\gamma_i(\vec{l},\chi)(\dobsj\gamma_j)^*(\vec{l''},\chi')\rangle
\end{align}
and subscripts $i$ and $j$ denote shears/overdensities/convergences taken from redshift bins $i$ and $j$. In Fourier space, the multiplicative adjustments to the shear become convolutions i.e. 
\begin{align}
(\kappa\gamma)(\vec{l}) &= \int \frac{d^2l'}{(2\pi)^2} \gamma(\vec{l'})\kappa(\vec{l}-\vec{l'})
\\
(\dobs\gamma)(\vec{l}) &= \int \frac{d^2l'}{(2\pi)^2} \gamma(\vec{l'}) \dobs(\vec{l}-\vec{l'}).
\label{eq:conv}
\end{align}
So 
\begin{align}
R_{ij} &=\int \frac{d^2l'}{(2\pi)^2} \langle\gamma_i(\vec{l'},\chi)\kappa_i(\vec{l}-\vec{l'},\chi) \gamma^*_j(\vec{l''},\chi')\rangle \\
R_{ji} &=\int \frac{d^2l'}{(2\pi)^2} \langle\gamma_i(\vec{l},\chi)\gamma_j^*(\vec{l'},\chi')\kappa_j(\vec{l''}-\vec{l'},\chi')\rangle \\
S_{ij} &= \int \frac{d^2l'}{(2\pi)^2} \langle\dobsi(\vec{l'},\chi)\gamma_i(\vec{l}-\vec{l'},\chi) \gamma^*_j(\vec{l''},\chi')\rangle \\
S_{ji} &=\int \frac{d^2l'}{(2\pi)^2} \langle\gamma_i(\vec{l},\chi)\gamma_j^*(\vec{l'},\chi')\dobsj^*(\vec{l''}-\vec{l'},\chi')\rangle. 
\end{align}
We use the following:
\begin{align}
\gamma_i(\vec{l}) &= e^{2i\phi_l}\kappa_i(\vec{l})\\
\kappa_i^*(\vec{l}) &= \kappa_i(-\vec{l})\\
\dobsi^*(\vec{l}) &= \dobsi(-\vec{l})
\end{align}
where $\phi_l$ is the angle made by $\vec{l}$ with the $x$-axis, to obtain
\begin{align}
R_{ij} &=\int \frac{d^2l'}{(2\pi)^2} e^{2i(\phi_{l'}-\phi_{l''})} \langle\kappa_i(\vec{l'},\chi)\kappa_i(\vec{l}-\vec{l'},\chi) \kappa_j(\vec{-l''},\chi')\rangle \label{eqn:rij_kkk}\\
R_{ji} &=\int \frac{d^2l'}{(2\pi)^2} e^{2i(\phi_{l}-\phi_{l'})} \langle\kappa_i(\vec{l},\chi)\kappa_j(\vec{-l'},\chi') \kappa_j(\vec{l'-l''},\chi')\rangle \\
S_{ij} &=\int \frac{d^2l'}{(2\pi)^2} e^{2i(\phi_{l'}-\phi_{l''})} \langle\kappa_i(\vec{l'},\chi)\dobsi(\vec{l}-\vec{l'},\chi) \kappa_j(\vec{-l''},\chi')\rangle \label{eqn:sij_dkk}\\
S_{ji} &=\int \frac{d^2l'}{(2\pi)^2} e^{2i(\phi_{l}-\phi_{l'})} \langle\kappa_i(\vec{l},\chi)\kappa_j(-\vec{l'},\chi') \dobsj(\vec{l'}-\vec{l''},\chi')\rangle \label{eqn:sji_dkk}.
\end{align}
We can write the reduced shear terms $R_{ij}$ and $R_{ji}$ in terms of the convergence bispectrum, $B_{(\kappa_1,\kappa_2,\kappa_3)}(\vec{l_1},\vec{l_2},\vec{l_3})$ defined as
\begin{equation}
\langle\kappa_i(\vec{l_1})\kappa_j(\vec{l_2}) \kappa_k,(\vec{l_3})\rangle = \
(2\pi)^2\delta_D(\vec{l_1}+\vec{l_2}+\vec{l_3})B_{(\kappa_1,\kappa_2,\kappa_3)}(\vec{l_1},\vec{l_2},\vec{l_3}).
\end{equation}
This can be related to the matter bispectrum using the Limber approximation
\begin{equation}
B_{(\kappa_1,\kappa_2,\kappa_3)}(\vec{l_1},\vec{l_2},\vec{l_3}) = \int \frac{d\chi}{\chi^4} W_1(\chi)W_2(\chi)W_3(\chi) B_{\delta}(\vec{k_1}, \vec{k_2}, \vec{k_3}; \chi)
\label{eqn:bk_limber}
\end{equation}
where $W_i(\chi)$ is the lensing kernel for redshift bin $i$ and $\vec{k_1}=\vec{l_1}/\chi$ etc. Note the $\delta_D(\vec{l_1}+\vec{l_2}+\vec{l_3})$ enforces a triangle configuration of the three vectors. So $R_{ij}$ and $R_{ji}$ become
\begin{align}
R_{ij} &=\int \frac{d^2l'}{(2\pi)^2} e^{2i(\phi_{l'}-\phi_{l''})} 
(2\pi)^2 B_{(\kappa_i,\kappa_i,\kappa_j)}(\vec{l'},\vec{l}-\vec{l'},\vec{-l})\\
R_{ji} &=\int \frac{d^2l'}{(2\pi)^2} e^{2i(\phi_{l}-\phi_{l'})} 
(2\pi)^2 B_{(\kappa_i,\kappa_j,\kappa_j)}(\vec{l},\vec{-l'},\vec{l'-l}).
\end{align}
We can write the LHS of \ref{eqn:fourier_start} as
\begin{equation}
\langle\gamma_i(\vec{l})\gamma_j^*(\vec{l''})\rangle = \
(2\pi)^2\delta^D(\vec{l}-\vec{l''})C_{ij}^{\kappa}(l),
\end{equation}
so the change in $C_{ij}^{\kappa}(l)$ due to reduced shear is
\begin{align}
\delta_{\mathrm{red}} C_{ij}^{\kappa}(l)&=[R_{ij}+R_{ji}]/(2\pi)^2 \notag\\
&=\int \frac{d^2l'}{(2\pi)^2} e^{2i(\phi_{l'}-\phi_{l})} \
B_{(\kappa_i,\kappa_i,\kappa_j)}(\vec{l'},\vec{l}-\vec{l'},\vec{-l}) \notag\\
&+ e^{2i(\phi_{l}-\phi_{l'})} 
B_{(\kappa_i,\kappa_j,\kappa_j)}(\vec{l},-\vec{l'},\vec{l'}-\vec{l}).\label{eq:red_long}
\end{align}
We can arrive at equation 13 of \citet{shapiro09} by taking the real part, assuming some symmetry properties of the convergence bispectrum ($B(\vec{l_1},\vec{l_2},\vec{l_3})=B(-\vec{l_1},-\vec{l_2},-\vec{l_3})$ and $B(\vec{l_1},\vec{l_2},\vec{l_3})=B(\vec{l_3},\vec{l_1},\vec{l_2})$) and defining the `2-redshift convergence bispectrum'
\begin{equation}
\begin{split}
B_{ij}(\vec{l_1},\vec{l_2},\vec{l_3})=\
\frac{1}{2}\int &\frac{d\chi}{\chi^4}W_i(\chi)W_j(\chi)[W_i(\chi)+W_j(\chi)]\\&B_{\delta}(\vec{k_1}, \vec{k_2}, \vec{k_3}; \chi),
\end{split}
\end{equation}
which in our notation is equal to 
\be
\frac{1}{2}[B_{(\kappa_i,\kappa_i,\kappa_j)}(\vec{l_1},\vec{l_2},\vec{l_3}) + B_{(\kappa_i,\kappa_j,\kappa_j)}(\vec{l_1},\vec{l_2},\vec{l_3})].
\ee
Substituting into \eqn{eq:red_long}
\begin{equation}
\delta_{\mathrm{red}} C_{ij}^{\kappa}(l)=\
2\int \frac{d^2l'}{(2\pi)^2} \textrm{cos}(2\phi_{l'}-2\phi_{l}) \
B_{ij}(\vec{l'},\vec{l}-\vec{l'},\vec{-l}).
\end{equation}
We now move on to the $S_{ij}$ and $S_{ji}$ terms. Various things can cause a galaxy overdensity $\delta_{obs}$, but we're concerned with ones that are correlated with the density field. These arise from two sources. The first and most obvious one is if the source galaxies trace the density field e.g. with some linear bias $b_g$
\begin{equation}
\dobsi(\vec{l},\chi)=N_i(\chi)b_g(\chi)\delta(\vec{l},\chi).
\end{equation}
Then we have 
\be
\begin{split}
S_{ij} =\int \frac{d^2l'}{(2\pi)^2} e^{2i(\phi_{l'}-\phi_{l''})} \langle&\kappa_i(\vec{l'},\chi)b_g(\chi)N_i(\chi)\\
&\delta(\vec{l}-\vec{l'},\chi) \kappa_j(\vec{-l''},\chi')\rangle.
\end{split}
\ee
In the Limber approximation (in which we assume density fluctuations at different radial distances are uncorrelated), this term goes to zero, by the following argument: $\kappa_i(\vec{l'},\chi)$ only depends on the density field for radial distances less than $\chi$, and so is uncorrelated with $\delta(\vec{l}-\vec{l'},\chi)$. $\kappa_j(\vec{-l''},\chi')$ gets contributions from density fluctuations all along the line of sight. Those produced by fluctuations at $\chi'!=\chi$ will be uncorrelated with $\delta(\vec{l}-\vec{l'},\chi)$, so for $\chi'!=\chi$, $\delta(\vec{l}-\vec{l'},\chi)$ is correlated with neither $\kappa_i(\vec{l'},\chi)$ or $\kappa_j(\vec{l'},\chi')$. The contribution to $\kappa_j(\vec{-l''},\chi')$ produced by fluctuations at $\chi'=\chi$ will be correlated with $\delta(\vec{l}-\vec{l'},\chi)$, but uncorrelated with $\kappa_i(\vec{l'},\chi)$. In both these cases, one of the variables in the 3-point correlator is uncorrelated with the other two, and since all variables have zero mean, the 3-point correlation is zero. Hence for $\dobs(\chi)$ satisfying $\langle\dobs(\chi)\delta(\chi')\rangle = \delta^D(\chi-\chi')\langle\dobs(\chi)\delta(\chi')\rangle$, $S_{ij} = S_{ji} = 0$. This is the source-lens clustering term which is zero in the Limber approximation (see \citealt{valegas14} for a treatment beyond the Limber approximation).

From \citet{schmidt09}, the lensing-bias produces an observed galaxy overdensity $\dobsi(\vec{\theta},\chi) = q_i\kappa_i(\vec{\theta},\chi)$ (to first order in $\kappa$), where $q$ is a constant that depends on the survey selection function. In this case, $S_{ij} = q_iR_{ij}$, and we get the same result as in the reduced-shear case, but for the $q_i$ prefactors
\begin{align}
\delta_{\text{lensing}}C_{ij}^{\kappa}(l) &=[q_iR_{ij}+q_jR_{ji}]/(2\pi)^2 \notag\\
&= \int \frac{d^2l'}{(2\pi)^2} \cos(2\phi_{l'}-2\phi_l) B^q_{ij}(\vec{l'},\vec{l}-\vec{l'},-\vec{l})\label{eq:lens_bias}
\end{align}
where
\be
\begin{split}
B^q_{ij}(\vec{l_1},\vec{l_2},\vec{l_3})=\
\frac{1}{2}\int \frac{d\chi}{\chi^4}&W_i(\chi)W_j(\chi)[q_iW_i(\chi)+q_jW_j(\chi)]\\
&B_{\delta}(\vec{l_1}/\chi, \vec{l_2}/\chi, \vec{l_3}/\chi; \chi).
\end{split}
\ee
This is a generalisation for tomography of the result of \citet{schmidt09}, who did not consider multiple redshift bins. \citet{schmidt09} show that the factor q has contributions from three effects. Let $f$, $r$ and $\vec{\theta}$ denote the observed flux, size and position of a galaxy, and $f_g$, $r_g$ and $\vec{\theta_g}$ the corresponding intrinsic (\textit{unlensed}) quantities. To first order in $\kappa$, the observed and intrinsic properties are related via
\be
\vec{\theta}=\vec{\theta_g}+\delta\vec{\theta},\, f=Af_g,\, r=\sqrt[]{A}r_g,\, d^2\vec{\theta}=Ad^2\vec{\theta_g}
\ee
where $A=1+2\kappa$. The first contribution to the observed galaxy overdensity comes from the change in the observed area element - a small patch of unlensed sky of area $\delta\theta^2$ has area $A\delta\theta^2$ due to lensing, and so $\dobs$ is reduced by a factor $A$. The second and third contributions come from the effect of magnification on the observed galaxies fluxes and sizes. In positive convergence regions, the magnification produces larger brighter galaxies, which are more likely to be detected and have well-measured shapes. \citet{schmidt09} show that the observed galaxy overdensity can be written as
\be
\dobs(\vec{\theta},\chi) = q\kappa(\vec{\theta},\chi) = (2\beta_f+\beta_r-2)\kappa(\vec{\theta},\chi)
\ee
where
\begin{align}
\beta_f \equiv \int dr \int df \frac{\partial\epsilon(f,r)}{\partial(\ln(f))}\Phi(f,r)\\
\beta_r \equiv \int dr \int df \frac{\partial\epsilon(f,r)}{\partial(\ln(r))}\Phi(f,r).
\end{align}
$\epsilon(f,r)$ is the selection function (i.e. accounts for the exclusion of faint, small galaxies) and $\Phi(f,r)$ is the true galaxy distribution in flux and size. These functions are normalised such that $\int df \int dr  \epsilon(f,r)\Phi(f,r)=1$. Hence if $\epsilon(f,r)$ is an increasing function of flux and size, $\beta_f$ and $\beta_r$ will be positive, since we'll observe a higher galaxy number density due to the magnification when $\kappa$ is positive.

\section*{Affiliations}
$^{1}$ Jodrell Bank Center for Astrophysics, School of Physics and Astronomy, University of Manchester, Oxford Road, Manchester, M13 9PL, UK\\
$^{2}$ Institut de F\'{\i}sica d'Altes Energies (IFAE), The Barcelona Institute of Science and Technology, Campus UAB, 08193 Bellaterra (Barcelona) Spain\\
$^{3}$ Department of Physics, ETH Zurich, Wolfgang-Pauli-Strasse 16, CH-8093 Zurich, Switzerland\\
$^{4}$ Fermi National Accelerator Laboratory, P. O. Box 500, Batavia, IL 60510, USA\\
$^{5}$ Kavli Institute for Cosmological Physics, University of Chicago, Chicago, IL 60637, USA\\
$^{6}$ Jet Propulsion Laboratory, California Institute of Technology, 4800 Oak Grove Dr., Pasadena, CA 91109, USA\\
$^{7}$ Department of Physics, University of Michigan, Ann Arbor, MI 48109, USA\\
$^{8}$ Department of Physics and Astronomy, University of Pennsylvania, Philadelphia, PA 19104, USA\\
$^{9}$ Department of Physics, Stanford University, 382 Via Pueblo Mall, Stanford, CA 94305, USA\\
$^{10}$ Kavli Institute for Particle Astrophysics \& Cosmology, P. O. Box 2450, Stanford University, Stanford, CA 94305, USA\\
$^{11}$ SLAC National Accelerator Laboratory, Menlo Park, CA 94025, USA\\
$^{12}$ Cerro Tololo Inter-American Observatory, National Optical Astronomy Observatory, Casilla 603, La Serena, Chile\\
$^{13}$ Department of Astrophysical Sciences, Princeton University, Peyton Hall, Princeton, NJ 08544, USA\\
$^{14}$ CNRS, UMR 7095, Institut d'Astrophysique de Paris, F-75014, Paris, France\\
$^{15}$ Department of Physics \& Astronomy, University College London, Gower Street, London, WC1E 6BT, UK\\
$^{16}$ Sorbonne Universit\'es, UPMC Univ Paris 06, UMR 7095, Institut d'Astrophysique de Paris, F-75014, Paris, France\\
$^{17}$ Laborat\'orio Interinstitucional de e-Astronomia - LIneA, Rua Gal. Jos\'e Cristino 77, Rio de Janeiro, RJ - 20921-400, Brazil\\
$^{18}$ Observat\'orio Nacional, Rua Gal. Jos\'e Cristino 77, Rio de Janeiro, RJ - 20921-400, Brazil\\
$^{19}$ Department of Astronomy, University of Illinois, 1002 W. Green Street, Urbana, IL 61801, USA\\
$^{20}$ National Center for Supercomputing Applications, 1205 West Clark St., Urbana, IL 61801, USA\\
$^{21}$ Institut de Ci\`encies de l'Espai, IEEC-CSIC, Campus UAB, Carrer de Can Magrans, s/n,  08193 Bellaterra, Barcelona, Spain\\
$^{22}$ Excellence Cluster Universe, Boltzmannstr.\ 2, 85748 Garching, Germany\\
$^{23}$ Faculty of Physics, Ludwig-Maximilians-Universit\"at, Scheinerstr. 1, 81679 Munich, Germany\\
$^{24}$ Department of Astronomy, University of Michigan, Ann Arbor, MI 48109, USA\\
$^{25}$ Department of Astronomy, University of California, Berkeley,  501 Campbell Hall, Berkeley, CA 94720, USA\\
$^{26}$ Lawrence Berkeley National Laboratory, 1 Cyclotron Road, Berkeley, CA 94720, USA\\
$^{27}$ Center for Cosmology and Astro-Particle Physics, The Ohio State University, Columbus, OH 43210, USA\\
$^{28}$ Department of Physics, The Ohio State University, Columbus, OH 43210, USA\\
$^{29}$ Australian Astronomical Observatory, North Ryde, NSW 2113, Australia\\
$^{30}$ Departamento de F\'{\i}sica Matem\'atica,  Instituto de F\'{\i}sica, Universidade de S\~ao Paulo,  CP 66318, CEP 05314-970, S\~ao Paulo, SP,  Brazil\\
$^{31}$ George P. and Cynthia Woods Mitchell Institute for Fundamental Physics and Astronomy, and Department of Physics and Astronomy, Texas A\&M University, College Station, TX 77843,  USA\\
$^{32}$ Instituci\'o Catalana de Recerca i Estudis Avan\c{c}ats, E-08010 Barcelona, Spain\\
$^{33}$ Department of Physics and Astronomy, Pevensey Building, University of Sussex, Brighton, BN1 9QH, UK\\
$^{34}$ Centro de Investigaciones Energ\'eticas, Medioambientales y Tecnol\'ogicas (CIEMAT), Madrid, Spain\\
$^{35}$ Brookhaven National Laboratory, Bldg 510, Upton, NY 11973, USA\\
$^{36}$ Institute of Cosmology \& Gravitation, University of Portsmouth, Portsmouth, PO1 3FX, UK\\
$^{37}$ Argonne National Laboratory, 9700 South Cass Avenue, Lemont, IL 60439, USA\\


\bsp	
\label{lastpage}
\end{document}